\documentclass[prl,aps,superscriptaddress,footinbib,twocolumn]{revtex4-2}
\usepackage[latin9]{inputenc}
\usepackage{color}
\usepackage{siunitx}
\usepackage{amsmath}
\usepackage{amssymb}
\usepackage{stmaryrd}
\usepackage{csquotes}
\usepackage{graphicx}
\usepackage[unicode=true,bookmarks=true,bookmarksnumbered=false,bookmarksopen=false,breaklinks=false,pdfborder={0 0 1},backref=false,colorlinks=true]
 {hyperref}
\hypersetup{
 linkcolor=magenta, urlcolor=blue, citecolor=blue, pdfstartview={FitH}, unicode=true}
 
\usepackage{todonotes}
\makeatletter
\usepackage{amsfonts}\usepackage{tabularx}\usepackage{dcolumn}\usepackage{bm}\usepackage{graphicx}\usepackage{epstopdf}
\usepackage{times}
\hypersetup{urlcolor=blue}

\def\ket#1{\left|#1\right\rangle}

\def\Bc{{\beta_{\text{C}}}}

\def\Hb{{\mathcal{H}_{\text{B}}}}

\makeatother

\begin{document}
\title{Probing many-body Bell correlation depth with superconducting qubits}
\affiliation{Zhejiang Key Laboratory of Micro-nano Quantum Chips and Quantum Control, School of Physics, Zhejiang University, Hangzhou 310027, China\\
$^2$ Center for Quantum Information, IIIS, Tsinghua University, Beijing 100084, China\\
$^{3}$ Instituut-Lorentz, Universiteit Leiden, P.O. Box 9506, 2300 RA Leiden, The Netherlands \\
$^{4}$ ZJU-Hangzhou Global Scientific and Technological Innovation Center, Hangzhou 310027, China\\
$^{5}$ Hefei National Laboratory, Hefei 230088, China\\
$^{6}$ Shanghai Qi Zhi Institute, Shanghai 200232, China}

\author{Ke Wang$^{1}$}\thanks{These authors contributed equally to this work.}
\author{Weikang Li$^{2,3}$}\thanks{These authors contributed equally to this work.}
\author{Shibo Xu$^{1}$}\thanks{These authors contributed equally to this work.}
\author{Mengyao Hu$^{3}$}
\author{Jiachen Chen$^{1}$}
\author{Yaozu Wu$^{1}$}
\author{Chuanyu Zhang$^{1}$}
\author{Feitong Jin$^{1}$}	
\author{Xuhao Zhu$^{1}$}
\author{Yu Gao$^{1}$}
\author{Ziqi Tan$^{1}$}
\author{Aosai Zhang$^{1}$}	
\author{Ning Wang$^{1}$}
\author{Yiren Zou$^{1}$}
\author{Tingting Li$^{1}$}
\author{Fanhao Shen$^{1}$}
\author{Jiarun Zhong$^{1}$}
\author{Zehang Bao$^{1}$}
\author{Zitian Zhu$^{1}$}
\author{Zixuan Song$^{4}$}
\author{Jinfeng Deng$^{1}$}
\author{Hang Dong$^{1}$}
\author{Xu Zhang$^{1}$}
\author{Pengfei Zhang$^{1,4}$}
\author{Wenjie Jiang$^{2}$}
\author{Zhide Lu$^{2}$}
\author{Zheng-Zhi Sun$^{2}$}
\author{Hekang Li$^{4}$}
\author{Qiujiang Guo$^{1,4,5}$}
\author{Zhen Wang$^{1,5}$}
\author{Patrick Emonts$^{3}$}
\author{Jordi Tura$^{3}$}
\author{Chao Song$^{1,5}$}
\email{chaosong@zju.edu.cn}
\author{H. Wang$^{1,4,5}$}
\email{hhwang@zju.edu.cn}
\author{Dong-Ling Deng$^{2,5,6}$}\email{dldeng@tsinghua.edu.cn}

\begin{abstract}
Quantum nonlocality describes a stronger form of quantum correlation than that of entanglement. 
It refutes Einstein's belief of local realism and is among the most distinctive and enigmatic features of quantum mechanics. 
It is a crucial resource for achieving quantum advantages in a variety of practical applications, ranging from cryptography and certified random number generation via self-testing to machine learning. 
Nevertheless, the detection of nonlocality, especially in quantum many-body systems, is notoriously challenging. 
Here, we report an experimental certification of genuine multipartite Bell correlations, which signal nonlocality in quantum many-body systems, up to ${24}$ qubits with a fully programmable superconducting quantum processor. 
In particular, we employ energy as a Bell correlation witness and variationally decrease the energy of a many-body system across a hierarchy of thresholds, below which an increasing Bell correlation depth can be certified from experimental data. 
As an illustrating example, we variationally prepare the low-energy state of a two-dimensional honeycomb model with ${73}$ qubits and certify its Bell correlations by measuring an energy that surpasses the corresponding classical bound with up to ${48}$ standard deviations. 
In addition, we variationally prepare a sequence of low-energy states
and certify their genuine multipartite Bell correlations up to ${24}$ qubits via energies measured efficiently by parity oscillation and multiple quantum coherence techniques. 
Our results establish a viable approach for preparing and certifying multipartite Bell correlations, which provide not only a finer benchmark beyond entanglement for quantum devices, but also a valuable guide towards exploiting multipartite Bell correlation in a wide spectrum of practical applications.
\end{abstract}

\maketitle

\begin{figure*}[t]
\center
\includegraphics[width=0.7\linewidth]{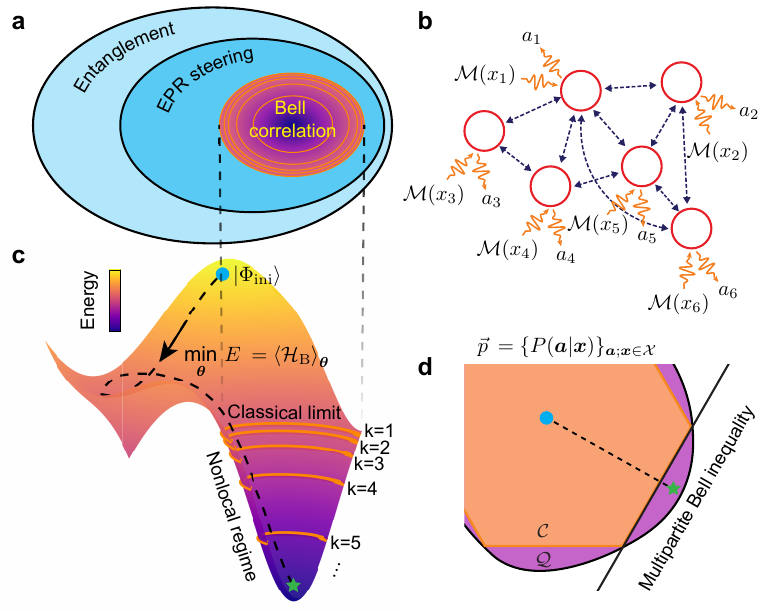}
\caption{\textbf{Many-body Bell correlation and the variational detection approach.}
\textbf{a}, The hierarchy of quantum correlations, starting from entanglement and Einstein-Podolsky-Rosen (EPR) steering, and finally to Bell correlation. Bell correlation is the strongest quantum correlation, in the sense that whenever it is detected then the other two types of correlation are guaranteed. In quantum many-body systems, Bell correlation can be further characterized by its depth (indicated by yellow contours), which quantifies the minimal number of particles sharing genuine nonlocal correlations.
\textbf{b}, A generic multipartite quantum system for detecting many-body Bell correlation, where site-resolved measurements $\mathcal{M}(
\mathbf{x})$ ($\mathbf{x}:=x_1,\dots,x_{N}$ with $x_i$ being the direction of measurement performed on the $i$th party) are performed on all partites and the corresponding outputs $\mathbf{a}:=a_1,\dots,a_{N}$ are collected.
The multipartite Bell inequality can be transformed into a many-body Hamiltonian by assigning each party the corresponding quantum observables. The detection of Bell correlation then becomes equivalent to finding low-energy states of the system with the energy exceeding the classical bound.
\textbf{c}, A schematic illustration of the essential idea of probing the many-body Bell correlation with a variational quantum circuit (VQC). 
Starting from the initial state $|\Phi_\text{ini}\rangle$, the VQC iteratively updates variational parameters $\boldsymbol{\theta}$ to minimize the energy of the many-body system $E=\langle\mathcal{H}_B\rangle_{\boldsymbol{\theta}}$, leading to violations of the corresponding Bell inequality and detection of the correlation depth. 
\textbf{d}, A sketch of the sets of correlations. For a given measurement configuration set $\mathcal{X}:=\{\mathbf{x}_1, \mathbf{x}_2, \dots, \mathbf{x}_K\}$, the probability distribution of all possible outcomes, denoted collectively by $ \vec{p}=	\{P(\mathbf{a}\vert \mathbf{x})\}_{\mathbf{a};\mathbf{x}\in\mathcal{X}}$, is confined in the polytope (light orange), which denotes 
the set of all possible classical correlations assuming local realism ($\mathcal{C}$). A facet of the polytope denotes a tight multipartite Bell inequality (black solid line), outside which certifies Bell correlations ($\mathcal{Q}$). 
}
\label{fig1}
\end{figure*}

\noindent
Quantum and classical physics differ fundamentally~\cite{Einstein1935Can}. 
Particles of a composite quantum system can exhibit correlations that are stronger than any classical theory permits. 
Several types of such correlations have been identified and experimentally verified: entanglement~\cite{Horodecki2009Quantum}, Einstein-Podolsky-Rosen (EPR) steering~\cite{Reid2009Colloquium}, and Bell nonlocality~\cite{Brunner2014Bell}. 
They serve as inequivalent resources for quantum technologies and form a hierarchical structure (Fig.~\ref{fig1}\textbf{a}), wherein the former is necessary but not enough to certify the latter~\cite{Wiseman2007Steering}. 
In particular, any quantum state that manifests Bell nonlocality, which can be experimentally validated by demonstrating violations of Bell inequalities~\cite{Bell1964Einstein,Clauser1969Proposed},  is necessarily entangled, but the opposite is \text{not} true: there exist quantum states that are entangled but admit a local-hidden-variable description, hence do not violate any Bell inequality and cannot show Bell nonlocality properties~\cite{Werner1989Quantum,Augusiak2015Entanglement}. 
Apart from this fundamental difference, Bell nonlocality (rather than entanglement) has been shown to be an indispensable resource for a wide range of practical applications, such as device-independent quantum key distribution~\cite{Acin2007DeviceIndependent,Vazirani2014Fully,Xu2020Secure}, certified random number generation via self-testing~\cite{Pironio2010Random}, and unconditional quantum computational advantages~\cite{Bravyi2018Quantum,LeGall2018Average,Gao2022Enhancing,Zhang2024Quantum}.  

Given its fundamental and practical importance, Bell nonlocality has been tested and confirmed repeatedly in various systems over the past 50 years, ranging from optical photons~\cite{Freedman1972Experimental,Aspect1981Experimental,Aspect1982Experimental,Weihs1998Violation,Giustina2015SignificantLoopholeFree,Shalm2015Strong,Li2018Test} 
and spins in nitrogen-vacancy centers~\cite{Hensen2015Loopholefree} to neural atoms~\cite{Hofmann2012Heralded,Schmied2016Bell,Engelsen2017Bell} and superconducting qubits~\cite{Storz2023Loopholefree}. 
The Nobel Prize in Physics 2022 was awarded jointly to Aspect, Clauser, and Zeilinger for their pioneering works in this direction~\cite{Hill2022Physics}. 
Yet, so far most existing experiments on Bell correlation have been focused on two- or few-body systems~\cite{Freedman1972Experimental,Aspect1981Experimental,Aspect1982Experimental,Weihs1998Violation,Hensen2015Loopholefree,Hofmann2012Heralded,Storz2023Loopholefree,Giustina2015SignificantLoopholeFree,Shalm2015Strong,Li2018Test}. 
Whereas entanglement has been extensively studied~\cite{Amico2008Entanglement,Horodecki2009Quantum} and hierarchies of genuine multipartite entanglement~\cite{Levi2013Hierarchies} have been reported with different experimental platforms, Bell nonlocality for quantum many-body systems remains much less explored.
Especially, the experimental observation of genuine multipartite nonlocality has mostly been limited to a handful of qubits~\cite{Mao2022Test,Cao2022Experimental}. 
Noteworthy previous experiments have indeed demonstrated the existence of Bell correlations through measuring certain collective observables in Bose-Einstein condensates~\cite{Schmied2016Bell} and spin-squeezed states with up to a half million rubidium atoms~\cite{Engelsen2017Bell}, but these systems lack individual addressing.
With recent progress in quantum computing, noisy intermediate-scale quantum devices with dozens to hundreds of individually addressable qubits are now available in laboratories~\cite{Preskill2018Quantum}.
This gives rise to a more and more pressing challenge of scaling up the demonstration of genuine Bell correlations to addressable many-body systems, so as to not only obtain a stronger quantumness benchmark beyond entanglement for these devices, but also lay down the foundation for their practical applications based on multipartite Bell correlations.

Here, we report for the first time an experimental certification of genuine multipartite Bell correlations up to $24$ particles on a fully programmable superconducting quantum processor. 
Our superconducting processor consists of up to $73$ qubits arranged on a two-dimensional ($2$D) lattice, with site-resolved controllability over all qubits and their nearest neighbor couplings. 
By upgrading the experimental setup~\cite{Xu2023nonAbelian} and optimizing the control and readout procedures,
we achieve parallel single (two)-qubit gates and simultaneous site-resolved qubit-state measurements with high precision, and the medians of the characteristic gate/measurement fidelities are benchmarked to be all above $99\%$.
This enables us to demonstrate Bell correlations in a fully-controllable quantum many-body system with unprecedented size and accuracy. 
To illustrate the validity and effectiveness of our approach,  we conduct three experiments.
We variationally prepare low-energy states of two spin Hamiltonians.
We certify Bell correlations for an XZ-type Hamiltonian on a $2$D honeycomb lattice with $73$ qubits by measuring an energy that surpasses the classical
bound with up to $48$  standard deviations. 
Additionally, we show Bell correlations in a one-dimensional XXZ-type Hamiltonian on $21$ qubits with up to $7$ standard deviations.
Furthermore, to detect genuine multipartite Bell correlations between more than two parties, we prepare a sequence of low-energy states and certify their Bell correlation depth,  which indicates the minimal number of parties sharing genuine nonlocal correlations in a composite system, by observing violations of the Svetlichny inequality~\cite{Svetlichny1987Distinguishing,Baccari2019Bell} on up to $24$ qubits.

\begin{figure*}[t]
\center
\includegraphics[width=1\linewidth]{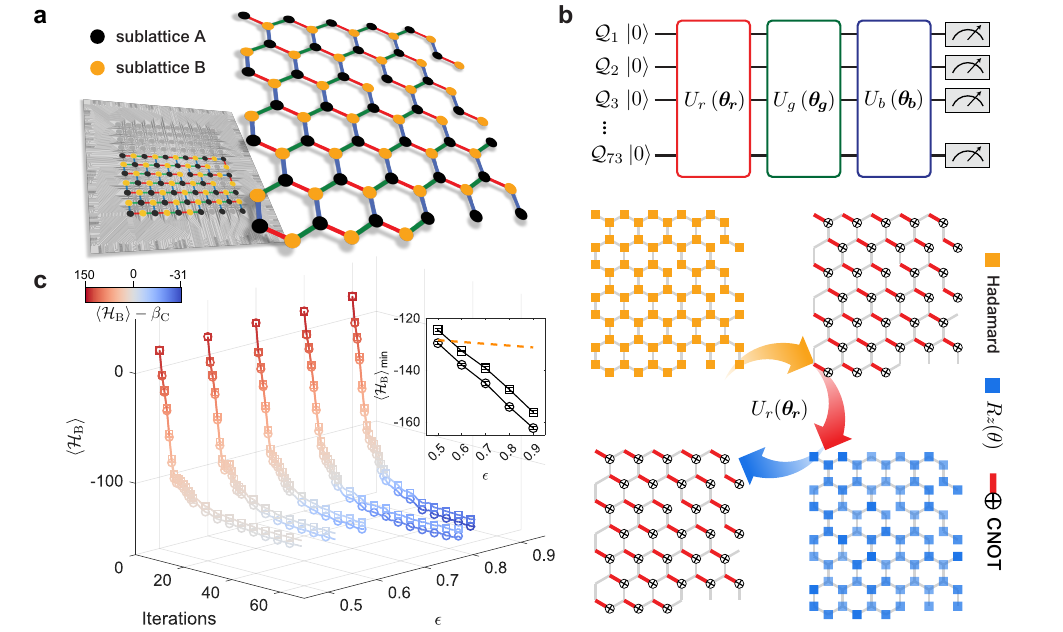}
\caption{\textbf{Detecting  Bell correlation with $\bold{73}$ transmon qubits on a $\bold{2}$D honeycomb lattice.}
\textbf{a}, Device and qubit topology. 
The superconducting chip contains a square lattice of 11$\times$11 transmon qubits, from which we select 73 qubits arranged in a brick-wall (honeycomb) lattice, which consists of two triangular sublattices (denoted here by A and B, respectively). 
Based on the explicit form of the many-body Bell inequality (\ref{BellHoneyComb}), the system Hamiltonian is constructed by assigning different weights to the coupling terms between the nearest neighbor qubits of different orientations as colored by red, green, and blue, respectively.
\textbf{b}, The variational quantum circuit designed for detecting Bell correlations.
This variational ansatz is composed of three blocks, each of which begins with a layer of Hadamard gates, followed by a variational entangler which contains a layer of independent $R_z(\theta)$ gates and is sandwiched by two layers of controlled-NOT (CNOT) gates.
For blocks with different colors, the CNOT gates are applied on qubit pairs connected by the links of the corresponding color as shown in \textbf{a}, with the circuit for the red block exemplified in the lower panel. 
\textbf{c}, 
Optimization trajectories for the XZ-type Hamiltonian $\mathcal{H}_B$ in Eq.~\eqref{XZ-type Hamiltonian 2D} with different coupling configurations characterized by $\epsilon$. As the variational parameters are updated iteratively, all energy values for different $\epsilon$ decrease below the classical bound $\beta_C$,
 indicating a violation of the corresponding Bell inequality \eqref{BellHoneyComb}. The 
 upper right inset shows the measured minimal energies and the classical bound (yellow dashed line) versus $\epsilon$ during the variational process.
 Here we show both the data with (circle) and without (square) readout corrections. Error bars represent the standard deviation with ten repetitions of experiments. 
}
\label{fig2}
\end{figure*}

\vspace{.5cm}
\noindent\textbf{\large{}General recipe and experimental setup}

\noindent
Although a number of Bell inequalities for quantum many-body systems have been discovered~\cite{Tura2014Detecting,Bancal2009Quantifying,Baccari2019Bell,Svetlichny1987Distinguishing,Mermin1990Extreme,Zukowski2002Bell,Brunner2014Bell,Frerot2023Probing}, the experimental demonstration of  Bell correlations by violating an inequality is exceedingly challenging, especially for the case of certifying genuine multipartite Bell correlations. 
There are at least two pronounced difficulties: (i) preparing highly-entangled quantum many-body states that violate these Bell inequalities with suitably chosen measurement settings, despite inevitable experimental noise in current quantum devices; (ii) some of these inequalities are constructed from correlation functions involving all parties and encompass an exponential number of correlators~\cite{Guhne2005Bell}.

For dealing with the first difficulty, we employ energy as a Bell correlation witness~\cite{Tura2017Energy} and introduce an approach based on variational quantum circuits (VQCs)~\cite{Cerezo2021Variational}. 
Our general recipe is illustrated in Fig.~\ref{fig1}. 
For a given Bell inequality written in the generic form $\mathcal{I}\geqslant\Bc$, where $\mathcal{I}$ is a linear function of correlators and $\Bc$ denotes the classical bound, we assign appropriate quantum measurement observables to each party involved in the inequality (Fig.~\ref{fig1}\textbf{b}). 
After choosing a set of measurements, we reinterpret $\mathcal{I}$ as a many-body Hamiltonian $\Hb$ and reduce the certification of Bell correlations to a task of finding quantum states with energies lower than the corresponding classical bounds (Methods). 
We start with an initialized variational quantum state $|\Phi(\boldsymbol{\theta})\rangle$, where $\boldsymbol{\theta}$ denotes collectively the variational parameters, and then iteratively update $\boldsymbol{\theta}$ to minimize the energy expectation $\langle\Phi(\boldsymbol{\theta})|\Hb|\Phi(\boldsymbol{\theta})\rangle$. 
The gradients required in updating $\boldsymbol{\theta}$ are obtained by exploiting the parameter shift rule~\cite{Mitarai2018Quantum} and measuring the difference between energies at two shifted parameters directly on the device (Methods).  
In general, the initial states would not show a violation of the corresponding Bell inequality. 
As we update $\boldsymbol{\theta}$ step-by-step, the energy decreases and becomes lower than the classical bound, indicating the presence of Bell correlations (Fig.~\ref{fig1}\textbf{c},\textbf{d}). 
Updating $\boldsymbol{\theta}$ further may decrease the energy further, leading to the certification of a larger Bell correlation depth (Fig.~\ref{fig1}\textbf{c}). 
To deal with the second difficulty, on the one hand we judiciously choose measurement settings to reduce the number of terms involved in $\Hb$ and on the other hand upgrade the experimental setup substantially. 
In particular, the measurement fidelity was improved from $94\%$~\cite{Xu2023nonAbelian} to $99\%$.

Our experiments are performed on a superconducting processor with up to 73 frequency-tunable transmon qubits arranged in a square lattice (Fig.~\ref{fig2}\textbf{a}). 
Each qubit has an on-chip control line for individual XY control (rotating around axes in the $xy$ plane of the Bloch sphere) and Z control (rotating around $z$ axis), which enable arbitrary single-qubit gates. 
Each qubit is also dispersively coupled to its own readout resonator to realize individual projective measurements. 
The nearest neighboring qubits are linked through a tunable coupler, which is also a transmon qubit with an individual on-chip control line, enabling the \textit{in-situ} tunability of the coupling strength. 
{Through optimizing the control and readout procedures, we push the median simultaneous single- (two-) qubit gate fidelity above $99.95\%$ ($99.4\%$) and the median measurement fidelity above $99.1\%$ (averaged for each qubit of readout error $|0\rangle$ and $|1\rangle$) on 73 qubits.}
During the implementation of VQCs, the couplings are turned off for parallel single-qubit gates and readout, and activated for applying two-qubit gates, which is essential for realizing high-fidelity quantum operations and measurements required in the detection of genuine multipartite Bell correlations.

We stress that the goal of our experiments is to probe and demonstrate Bell correlations (especially genuine multipartite Bell correlations) in a many-body system, rather than disprove the local realism nature of the world as in traditional Bell tests~\cite{Freedman1972Experimental,Aspect1981Experimental,Aspect1982Experimental,Weihs1998Violation,Hensen2015Loopholefree,Hofmann2012Heralded,Storz2023Loopholefree,Giustina2015SignificantLoopholeFree,Shalm2015Strong,Li2018Test}. 
Hence, we do not require the used qubits to be space-like separated from each other to close the locality loophole. In fact, the single-qubit measurement duration in our experiments (about $\SI{2}{\micro\second}$) is much longer than the time for light to transfer between transmon qubits. 
In addition, when reducing the number of measurement settings in the experiment, we assume the validity of quantum mechanics, which is a natural and commonly used assumption in detected Bell correlations in many-body systems~\cite{Schmied2016Bell,Engelsen2017Bell}.

\begin{figure*}[t]
\center
\includegraphics[width=1\linewidth]{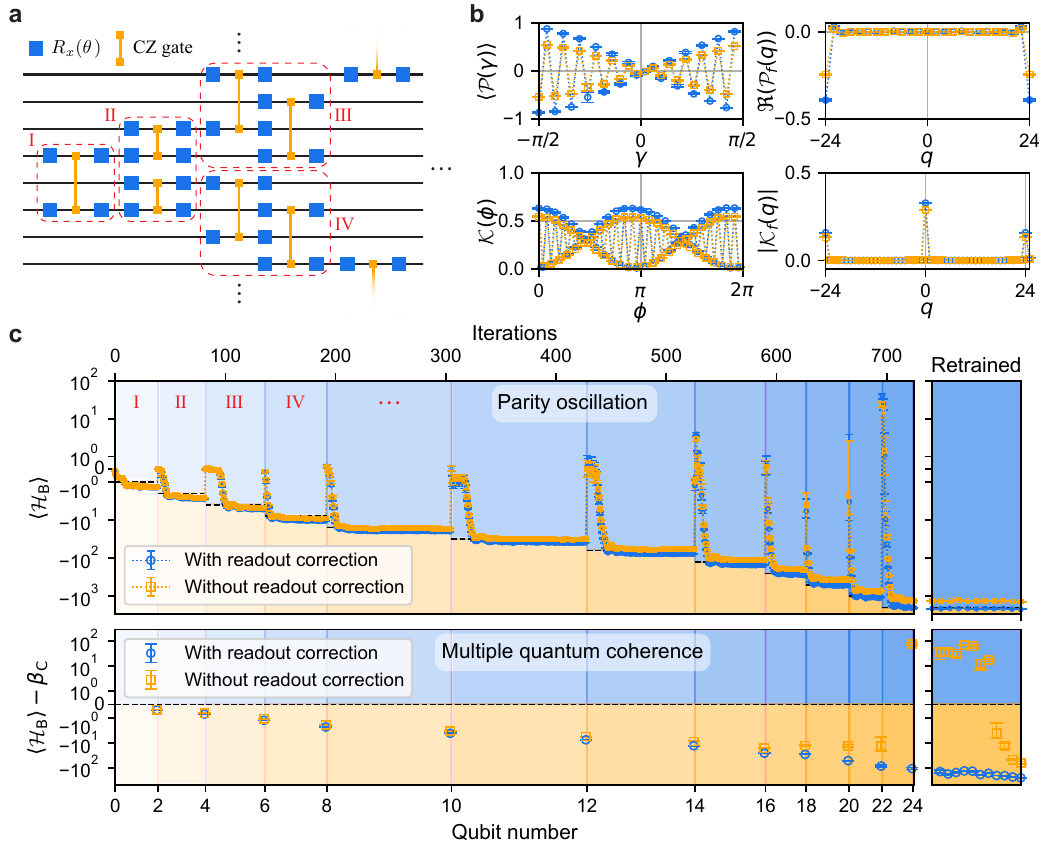}
\caption{\textbf{Detection of Bell correlation depth.}
\textbf{a}, The variational quantum circuit with a hierarchical architecture designed for probing Bell correlation depth.
The circuit is divided into many sub-circuits labelled I, II, III, etc., which are trained sequentially to avoid the problem of barren plateaus. 
Starting with two qubits, we train sub-circuit I to minimize the energy of the two-qubit system, following which we expand the system by involving two more qubits and train sub-circuit II to minimize the energy of the four-qubit system.
The procedure goes on until all $24$ qubits are involved, during which the detected Bell correlation depth increases step by step.
\textbf{b}, Measured parity expectation $\left\langle\mathcal{P}(\gamma)\right\rangle$ and multiple quantum coherence (MQC) $\mathcal{K}(\phi)$ data for the $24$-qubit Greenberger-Horne-Zeilinger (GHZ) state with (blue circles) and without (yellow squares) readout correction, alongside their corresponding Fourier spectrum ($\mathcal{P}_f(q)$ and $\mathcal{K}_f(q)$) from which the energy of the system $\langle \mathcal{H}_B\rangle$ with $\mathcal{H}_B$ shown in Eq.~\eqref{eq:k-nonlocal_Hamiltonian_ev} can be obtained. 
Error bars denote the standard deviation from five repetitions of the experiment.
\textbf{c}, Optimization trajectory. 
During the training procedure, we assess the loss function -- the energy of the system -- based on parity measurement. 
The whole training procedure is separated into $13$ phases, with the first $12$ phases corresponding to the hierarchical training procedure and the last phase retraining the whole circuit again. 
The energy values with (blue circles) and without (orange squares) readout correction are presented for each phase, with error bars denoting the standard deviation from five repetitions of the experiment. 
Dashed lines indicate the  $k$-nonlocal bounds (lower than which guarantees Bell correlation depth $k+1$) for the corresponding phase. The lower panel shows the energy differences relative to the $k$-nonlocal bounds, with the energies measured by the MQC method at the end of each training phase, which are less affected by readout errors ({Methods and Supplementary Sec.~II.C}). }
\label{fig3}
\end{figure*}

\vspace{.5cm}
\noindent\textbf{\large{}An example on the honeycomb lattice}

\noindent
As an illustrative example, we first consider a multipartite Bell inequality defined on a $2$D honeycomb lattice.  
We label the two sublattices of the honeycomb lattice by $A$ and $B$, and the links between them by different colors $\tau \in \{\text{r}, \text{b}, \text{g}\}$, as shown in Fig.~\ref{fig2}\textbf{a}. 
We assign two choices of measurements $x,y \in \{0,1\}$ with outcomes $a,b\in \{-1,+1\}$ for particles living on sublattices $A$ and $B$, respectively.
The multipartite Bell inequality  reads~\cite{Deng2018Machine}:
\begin{eqnarray}\label{BellHoneyComb}
\mathcal{I} = \sum_{\tau-\text{link}} J_\tau (\epsilon)\sum_{x,y}(-1)^{x\cdot y}\langle A_xB_y \rangle_{\tau} 
 \geq \Bc,
\end{eqnarray}
where $\Bc = -2\sum_{\tau} J_\tau (\epsilon)$, $\langle A_xB_y\rangle =\sum_{a,b} abP(ab \vert xy)$, and  $J_\tau (\epsilon) = 1+\epsilon$ and $\frac{1-\epsilon}{2}$ for $\tau =\text{r}$ and $\tau=\text{b},\text{g}$, respectively. 
By choosing proper measurement settings (Methods), the corresponding Bell operator reduces to an XZ-type Hamiltonian:
\begin{eqnarray}\label{XZ-type Hamiltonian 2D}
\Hb=\sum_{\tau-\text{link}} J_\tau (\epsilon)(\sigma_x^{A}\sigma_x^{B}+\sigma_z^{A}\sigma^{B}_z)_{\tau},    
\end{eqnarray}
where $\sigma_x$ and $\sigma_z$ are Pauli operators acting on the two qubits at the ends of the link $\tau$ and the superscript denotes the sublattice they belong to.
For this particular setting, the quantum violation of the inequality~\eqref{BellHoneyComb} corresponds to a system energy $\langle\Hb\rangle$ that is lower than the classical bound $\Bc$. 
In particular, the maximal violation corresponds to the ground-state energy. 
Yet, $\Hb$ is a $2$D many-body Hamiltonian that is not exactly solvable. 
As a result, determining its ground-state energy is classically intractable in general. 
In our approach, we exploit the VQC ansatz to tackle this challenge and detect many-body Bell correlations. 
We design a parameterized quantum circuit ansatz with three blocks, each containing single-qubit rotation and two-qubit controlled-NOT (CNOT) gates, as shown in Fig.~\ref{fig2}\textbf{b}. 
During the optimization process, we use the parameter shift rule~\cite{Mitarai2018Quantum} to obtain the gradients (i.e., the derivatives of the model Hamiltonian's energy with respect to the variational parameters) by measuring corresponding observables.

In Fig.~\ref{fig2}\textbf{c}, we plot our experimental results for detecting multipartite Bell correlation with the Bell inequality~\eqref{BellHoneyComb} and five different coupling strengths $\epsilon = 0.5, 0.6,\dots,\text{and}\ 0.9$. We first initialize the system to a product state $|0\rangle^{\otimes 73}$ and then let it evolve under the parameterized quantum circuit ansatz.
At the beginning, the variational parameters are arbitrarily chosen to be one and the measured energy (expectation value $\langle \Hb\rangle$) is larger than the classical bound $\Bc$, hence no quantum violation and no  Bell correlation is detected.
We then update the variational parameters with the measured gradients to decrease the energy step by step.
As shown in Fig.~\ref{fig2}\textbf{c}, the energy decreases gradually as the number of iterations increases.
It goes below the classical bound after a few iterations, leading to quantum violations and the detection of Bell correlations.
Taking the curve for $\epsilon = 0.9$ as an example, we observe that the corresponding measured energy goes below the classical bound after $14$ iterations and converges to its minimal value of $-156.29$, which is lower than the classical bound {(-131.3)} by 48 standard deviations and unambiguously signatures Bell correlations.

We note that in general the experimentally measured minimal energies might have deviations from their theoretical predictions which are hard to obtain classically.
This is due to experimental imperfections (such as gate errors and decoherence) and limitations on the expressivity of the variational ansatz.
We also stress two crucial differences between this experiment and the previous ones with rubidium atoms~\cite{Schmied2016Bell,Engelsen2017Bell}: (i) the previous experiments focused on ensembles of atoms without individual addressing and detected Bell correlations through measuring collective observables, whereas in our experiment the $73$ transmon qubits used can be individually manipulated with high precision. 
The former requires an underlying assumption that the collective observable is a sum of individual observables, which we do not need to make here.
(ii) the previous experiments rely on carefully designed Bell correlation witnesses suitable for spin-squeezed states and thus lack general applicability to other Bell inequalities and more general quantum states. 
In contrast, the variational approach exploited in the current experiment is generically applicable to any Bell inequality, independent of its particular structure~\cite{Kokail2019Selfverifying}. 
Indeed, the general applicability of our approach is further manifested in two additional experiments, one for detecting Bell correlations for a $1$D chain system (Extended Data Fig.~1) and the other for detecting correlation depth (Fig.~\ref{fig3}).

\vspace{.5cm}
\noindent\textbf{\large{}Detecting Bell correlation depth}

\noindent
In the above discussion, Bell correlations are detected once the energy is lower than the classical bound. 
Yet, in this scenario, only the bipartite Bell correlation is certified from the experimental results. 
As an example, a single singlet in an otherwise classical state will register as a nonlocal state.
This is not quite satisfactory.
In this section, we turn to a more challenging problem of detecting genuine multipartite Bell correlation. 
We consider a system consisting of $N$ qubits with each assigned two possible dichotomic measurements (labeled by $x_i$, $i \in \{1,2,\dots, N\}$). 
We exploit the following Svetlichny inequality~\cite{Svetlichny1987Distinguishing,Baccari2019Bell}: 
\begin{eqnarray}
    \label{eq:Sve_ineq_N22}
    \mathcal{I}_{N}^{\text{Sv}}&=&2^{-\frac{N}{2}} \left[ \sum_{\mathbf{x} \vert s = 0}(-1)^{\frac{s}{2}} \langle \mathbf{A}\rangle_{\mathbf{x}}+ \sum_{\mathbf{x} \vert s = 1}(-1)^{\frac{{s-1}}{2}} \langle \mathbf{A}\rangle_{\mathbf{x}}\right],\nonumber\\
   &\geq& -2^{\left(N-\left\lceil\frac{N}{k}\right\rceil\right) / 2},\nonumber
\end{eqnarray}
where $\langle \mathbf{A}\rangle_{\mathbf{x}}=\langle A_{1,x_1}\dots A_{N,x_N}\rangle$ denotes the $N$-partite correlator, and $s$ is the parity of $\sum_{i}x_i$.
This inequality serves as a witness to a genuine multipartite nonlocality. 
Any violation of this inequality would guarantee that the Bell nonlocality depth is at least $k+1$, i.e., there are at least $k+1$ parties sharing genuine nonlocal correlations. 
Yet, there are two apparent challenges in the experimental detection of genuine multipartite Bell correlation with this inequality. 
First, the number of correlators involved in the Svetlichny inequality grows exponentially with the system size. 
Second, each correlator is a $N$-partite correlator, which renders its measurement exceedingly difficult as $N$ increases. 
To overcome the first challenge, we choose a particular set of measurement settings so that the corresponding Bell operator of $\mathcal{I}_{N}^{\text{Sv}}$ reduces to
\begin{eqnarray}\label{eq:Sve_ham_N22}
    \Hb(N) =2^{\frac{N-1}{2}}[(|0\rangle\langle 1|)^{\otimes N}+(|1\rangle\langle0|)^{\otimes N}].\label{eq:k-nonlocal_Hamiltonian_ev}
\end{eqnarray}
$\Hb(N)$ contains only two terms, yet it still requires $N$-partite coincidence measurements. 
To overcome this challenge, we substantially upgraded the experimental setup and optimized the control and readout procedures, achieving the state-of-the-art average measurement fidelity of {$99.1\%$ (Methods and Supplementary)}. 
In addition, we exploit both parity oscillation~\cite{Song2019Generation} and multiple quantum coherence~\cite{MQC2020} techniques to measure multipartite correlations in an efficient fashion.

With the upgraded experimental setup and the strategies discussed above, we still face one more subtle challenge that prevails in VQC algorithms---the barren plateau problem (i.e., the gradients vanish exponentially with the system size~\cite{McClean2018Barren}). 
Indeed, as the system size and circuit depth increase, we find that the measured gradients decrease and would be completely washed out by experimental noise. 
Consequently, the training of the variational circuits eventually fails and we would not be able to detect a correlation depth larger than eight in our experiment. 
To deal with this challenge, we build up the variational circuit step by step and design a layer-wise training strategy~\cite{Bengio2006Greedy}, as illustrated in Fig.~\ref{fig3}. 
The variational quantum circuit starts with two qubits and the algorithm is applied to prepare a low-energy state of $\Hb(2)$. 
At each step, we add two more qubits and one more layer. 
We fix the parameters in the previous layers and variationally update the parameters of the new layer with measured gradients corresponding to the Hamiltonian $\Hb(N')$. 
With this approach and additional retraining of the whole variational circuit, we are able to overcome the barren plateau problem, and prepare and detect genuine multipartite nonlocality up to $24$ qubits. 

In Fig.~\ref{fig3}\textbf{b}, we benchmark the performance of the parity oscillation and multiple quantum coherence methods exploited in our experiment. 
We prepare the $24$-qubit Greenberger-Horne-Zeilinger (GHZ) state and measure the parity expectation
$\left\langle\mathcal{P}(\gamma)\right\rangle$ 
and multiple quantum coherence
$\mathcal{K}(\phi)$, where $\gamma$ and $\phi$ correspond to the single-qubit rotation angles in the corresponding measurement circuits (Methods). 
We find that the experimental data agree well with the theoretical prediction. 
In Fig.~\ref{fig3}\textbf{c}, we plot the optimization trajectory for the whole process. 
From this figure,  genuine $k$-correlation depths are observed with $k$ spanning from $2$ to $24$ in different learning phases for data both with and without readout corrections.

\vspace{.5cm}
\noindent\textbf{\large{}Conclusion and outlook}

\noindent
In summary, we have introduced a variational-quantum-circuit approach for probing Bell correlations in many-body systems, and experimentally observed genuine multipartite Bell correlations up to $24$ qubits with a programmable superconducting quantum processor. 
In contrast to previously reported detection of Bell correlations in ensembles of atoms without individual addressing, our experiments are performed on a fully programmable superconducting quantum processor with state-of-the-art gate and measurement fidelities. 
In addition, the variational approach exploited in our experiment is generically applicable to a broad family of Bell inequalities, independent of its specific structure and the qubit connection geometry of the quantum device.

Our results not only establish a practical approach to the preparation and certification of genuine multipartite Bell correlations with NISQ devices, but also provide a stronger benchmark beyond entanglement for testing how quantum these devices are. 
The controllability and scalability demonstrated in our experiment open up several new avenues for future studies. 
In particular, it would be interesting and important to demonstrate exponentially increasing Bell correlations with our current quantum processor~\cite{Bonsel2024Generating}. 
Another interesting direction is to explore VQCs with certain symmetries~\cite{Larocca2022GroupInvariant,Meyer2023Exploiting,Zheng2023Speeding} or use different optimization strategies, such as dissipative cooling~\cite{Mi2024Stable} and quantum approximate optimization algorithms~\cite{Farhi2014Quantum,Zhou2020Quantum,Dupont2023Quantumenhanced}, to mitigate the barren plateau problem and enhance the effectiveness of our protocol. 
It would also be interesting to extend our approach to alternative definitions of nonlocality depth based on local observables and shared randomness~\cite{Cao2022Experimental}.
From the perspective of practical applications, an experimental demonstration of quantum learning advantages~\cite{Gao2022Enhancing,Zhang2024Quantum} based on Bell correlations would mark a milestone towards future utilization of quantum technologies in artificial intelligence.

\vspace{.5cm}
\noindent\textbf{\large{}Methods}{\large\par}

\noindent\textbf{Bell inequality on a honeycomb lattice}

\noindent 
The Bell inequality introduced in Eq.~\eqref{BellHoneyComb} is composed of local terms shared by two connected parties.
The full expression of a local term is a CHSH inequality~\cite{Clauser1969Proposed}
\begin{equation}
    \mathcal{I}_{\text{local}} = \langle A_0B_0 \rangle + \langle A_0B_1 \rangle + \langle A_1B_0 \rangle - \langle A_1B_1 \rangle \geq -2, \nonumber
\end{equation}
where the notations follow the main text and we have generalized this inequality to a honeycomb lattice.

To create a map from the inequality defined on a honeycomb lattice to a Hamiltonian, we choose measurements 
$A_0 = \sigma_z\text{, }A_1 = \sigma_x$
for all parties belonging to sublattice $A$, and 
$B_0 = \frac{1}{\sqrt{2}}(\sigma_z + \sigma_x)\text{, }B_1 = \frac{1}{\sqrt{2}}(\sigma_z - \sigma_x)$
for all parties belonging to sublattice $B$.
Summing over all parties yields the Hamiltonian
$
\Hb=\sum_{\tau-\text{link}} J_\tau (\epsilon)(\sigma_x^A \sigma_x^B+\sigma_z^A \sigma_z^B)_{\tau}
$
studied in the main text.

\vspace{.5cm}
\noindent\textbf{Bell inequality and correlations in a $\mathbf{1}$D chain}

\noindent 
The $1$D chain system we consider in this work starts from Gisin's elegant inequality~\cite{Gisin2009Bell} (Supplementary), which involves two parties $A$ and $B$ with four and three dichotomic measurements, respectively. 
These measurements produce outcomes $\pm 1$.
Here, we use a modified version of Gisin's elegant inequality
\begin{equation}
\label{eq:XXZ_delta}
\begin{aligned}
   \mathcal{I}_{\Delta}&=\langle A_0 B_0 \rangle + \langle A_1 B_0 \rangle - \langle A_2 B_0 \rangle - \langle A_3 B_0 \rangle
   \\&+\langle A_0 B_1 \rangle - \langle A_1 B_1 \rangle + \langle A_2 B_1 \rangle - \langle A_3 B_1 \rangle
   \\&+\Delta(\langle A_0 B_2 \rangle - \langle A_1 B_2 \rangle - \langle A_2 B_2 \rangle + \langle A_3 B_2 \rangle)
   \\&\geqslant -(2|\Delta|+|\Delta+2|+|\Delta-2|),
\end{aligned}
\end{equation}
where $\Delta$ is a real number~\cite{Tura2017Energy}.

Given an odd number $N \geq 3$, we define a $1$D-chain model involving $N$ parties.
For each pair of neighboring parties, we assign the inequality described in Eq.~\eqref{eq:XXZ_delta}, where the odd sites play the role of party $A$.
The overall inequality results from a weighted sum of two-party inequalities
$
\mathcal{I}_{\text G}:=\sum_{i=1}^{N-1}(1-(-1)^i\epsilon)\mathcal{I}_{\Delta}^{(i,i+1)},
$
where $\mathcal{I}_{\Delta}^{(i,i+1)}$ denotes the inequality $\mathcal{I}_{\Delta}$ acting on sites indexed by $i$ and $i+1$. 
The classical bound $\Bc$ is $\Bc=-(N-1)(2|\Delta|+|\Delta+2|+|\Delta-2|)$.

The mapping from this inequality to the Hamiltonian studied in this work can be achieved by assigning specified measurement settings (Supplementary)~\cite{Gisin2009Bell,Tura2017Energy}.
In this way, we obtain an XXZ-like Hamiltonian on a $1$D chain
\begin{equation}
\label{eq:XXZ_H}
\Hb=\sum_{i=1}^{N-1} J_i (\epsilon)\left(\sigma_x^{(i)} \sigma_x^{(i+1)}+\sigma_y^{(i)} \sigma_y^{(i+1)}+\Delta \sigma_z^{(i)} \sigma_z^{(i+1)}\right),\nonumber
\end{equation}
where $J_i (\epsilon):=\frac{4}{\sqrt{3}}\left[1-(-1)^i \epsilon\right]$ is the coupling strength that varies concerning the site's parity, and $\Delta$ is a real number.

For the experiment demonstrated on the superconducting processor, we start with a product state, where the goal is to variationally prepare a low-energy state for the above Hamiltonian with an energy below the classical bound.
In this experiment, we set $\Delta=2$ and $\epsilon=0.95$. 
The VQC consists of single-qubit rotation gates and two-qubit controlled-Z (CZ) gates.
We record the observed energy at each iteration during the training process, as plotted in Extended Data Fig.~\ref{ex_fig:XXZ}.
After $34$ iterations of optimization, the energy of the prepared state reaches around $\langle \Hb\rangle=-163.7$, achieving $7$ standard deviations lower than the classical bound {-160}.
Compared with the $2$D experiment implemented on the honeycomb lattice, this experiment achieves the violation with a lower confidence level.
This may be caused by the smaller relative gap between the ground-state energy and the classical bound of this model in contrast to the $2$D one.

\vspace{.5cm}
\noindent\textbf{Detecting the Bell correlation depth}

\noindent The Mermin and Svetlichny Bell inequalities are used to detect the Bell correlation depth for the multipartite scenario in this work~\cite{Mermin1990Extreme,Svetlichny1987Distinguishing,Barreiro2013Demonstration, Baccari2019Bell, Bancal2009Quantifying}. 
While the inequalities are in principle able to detect the nonlocality depth, we refer to it as Bell correlation depth here due to open loopholes like the locality loophole.
We consider the $(N,2,2)$ scenario, i.e., there are two measurements $x_i \in \{0,1\}$ and two outcomes $a_i \in \{0,1\}$ for parties $i\in \{1,\dots,N\}$. 
The Hamiltonian introduced in Eq.~\eqref{eq:Sve_ham_N22} coincides with the Bell operators of these two inequalities of $N$ parties. 
Here we provide the Svetlichny Bell expression and assign the two measurements
\begin{equation}
    \begin{aligned}
          &A_{i,0}=\cos\left(\phi_1\right)\sigma_x+\sin\left(\phi_1\right)\sigma_y, \\
          &A_{i,1}=\cos\left(\phi_2\right)\sigma_x+\sin\left(\phi_2\right)\sigma_y
    \end{aligned}
    \nonumber
\end{equation} 
for each party $i \in \{1,\dots,N\}$, where $\phi_1=-\frac{\pi}{4N}$ and $\phi_2=\frac{(2N-1)\pi}{4N}$ (Supplementary). 
By choosing measurement settings (Supplementary), we obtain the Bell operator in the form of $\Hb(N) =2^{\frac{N-1}{2}}[(|0\rangle\langle 1|)^{\otimes N}+(|1\rangle\langle0|)^{\otimes N}]$.

\vspace{.5cm}
\noindent\textbf{Quantum circuit optimization}

\noindent
We transform the problem of finding Bell violations into searching for low-energy states of the corresponding Hamiltonian with VQCs. 
The optimization strategy for this framework starts with a product state followed by a parameterized quantum circuit
$
|\psi(\boldsymbol{\theta})\rangle=U(\boldsymbol{\theta})|\mathbf{0}\rangle
$,
and we choose the energy as the loss function, i.e., 
$
E(\boldsymbol{\theta})=\langle\psi(\boldsymbol{\theta})|\Hb| \psi(\boldsymbol{\theta})\rangle
$, that we aim to minimize.

In our experiments, we apply gradient-based methods to update the parameters $\boldsymbol{\theta}$ within the quantum circuit.
To calculate the gradients of the energy with respect to the parameters, we apply the parameter shift rule~\cite{Mitarai2018Quantum}.
In this work, the variational quantum circuit consists of parameterized single-qubit gates and two-qubit CNOT/CZ gates.
Since the variational parameters are encoded in the rotation angles in the form of $R(\theta_k)=e^{-i\theta_kP/2}$ with $P$ being a single Pauli operator $P \in \{\sigma_x,\sigma_y,\sigma_z\}$, and $\theta_k$ being the rotation angle,
the corresponding derivative can be analytically computed according to two expectation values
\begin{equation}
    \frac{\partial E(\bm{\theta})}{\partial\theta_k}=\frac{1}{2}(E(\bm{\theta}\backslash\theta_k,\theta_k+\pi/2)-E(\bm{\theta}\backslash\theta_k,\theta_k-\pi/2)),
\end{equation}
where $E(\bm{\theta}\backslash\theta_k,\theta_k\pm\pi/2)$ denotes the loss function with  $\theta_k$ replaced by $\theta_k\pm\pi/2$ and the other parameters unchanged.
After obtaining the gradients $\nabla_{\boldsymbol{\theta}} E\left(\boldsymbol{\theta}_n\right)$ from experiments, the parameters 
are updated with the Adam optimizer.

\vspace{.5cm}
\noindent\textbf{Many-body quantum correlation}

\noindent
The detection of Bell correlation depth relies on the ability of measuring many-body quantum correlation, which, in this work, is reduced to the detection of the matrix element $\mathcal{C}\equiv|0\rangle\langle 1|^{\otimes N}$.
The expectation value of the multipartite Hamiltonian $\Hb(N)$ can be obtained by $\langle\Hb(N)\rangle=2^{\frac{N-1}{2}}(\langle\mathcal{C}\rangle+\langle\mathcal{C}\rangle^*)$.
In practice, we can measure $\langle\mathcal{C}\rangle$ efficiently with either parity oscillation~\cite{20q_ghz} or multiple quantum coherence (MQC)~\cite{MQC2020} techniques, with the corresponding quantum circuits illustrated in Extended data Fig.~\ref{ex_fig:circuit}.

In the parity oscillation method, the collective spin operators $\mathcal{P}(\gamma)=\otimes_{j=1}^N\left[\sin(\gamma) \sigma_{y, j}+\cos(\gamma) \sigma_{x, j}\right]$ are measured for a given set of $\gamma\in[-\pi/2, \pi/2]$, and the matrix element can be obtained by Fourier transforming the resulting signal at the fixed frequency of $N$ as $\langle\mathcal{C}\rangle(N) = \mathcal{N}_s^{-1}\sum_{\gamma}e^{-iN\gamma}\langle\mathcal{P}(\gamma)\rangle$, where $\mathcal{N}_s$ ($=N+1$ in our experiment) denotes the sampling number of $\gamma$. 
The measurement with the MQC method involves more quantum operations: 
after preparing the quantum state with the variational unitary $U(\boldsymbol{\theta})$,  we apply a layer of refocusing $R_x(\pi)$ gates, a layer of $R_z(\phi)$ gates, and a reversed state preparation unitary $U^\dagger(\boldsymbol{\theta})$ before the projective measurement of the $|0\rangle^{\otimes N}$ state. 
The absolute value of the matrix element can be obtained based on the measured probability $\mathcal{K}(\phi)$ as $|\langle\mathcal{C}\rangle(N)| = \sqrt{\mathcal{N}_s^{-1}\left|\sum_\phi e^{iN\phi} \mathcal{K}(\phi)\right|}$, where $\mathcal{N}_s$ ($=2N+2$ in our experiment) denotes the sampling number of $\phi$ (Supplementary II.C). 
Combined with the phase information from the parity oscillation method, we can obtain $\langle\Hb(N)\rangle$.

\vspace{.5cm}
\noindent\textbf{Readout error mitigation}

\noindent
Readout is often the most error-prone elementary operation on superconducting quantum processors, especially when carried out simultaneously for many qubits.
It is crucial to mitigate the readout errors for the detection of many-body nonlocality correlations.
In our experiment, we realize median assignment errors of $0.008$ and $0.009$ for the simultaneous readout of $|0\rangle$ and $|1\rangle$ states on $73$ qubits, respectively.
This is achieved by implementing the excited state promotion technique~\cite{Elder_2020, Jurcevic_2021}, which essentially leverages the higher energy levels of the transmon qubit to reduce the state relaxation error and improve the signal-to-noise ratio of the readout procedure.

We further mitigate the impact of readout errors by adopting the MQC technique instead of the parity oscillation method in determining the Bell correlation depths. 
Specifically, assuming a symmetry readout assignment error $e$ of the $|0\rangle$ and $|1\rangle$ states, we have $\langle\mathcal{C}\rangle \approx \left(1-e\right)^{N/2}\langle\mathcal{C}\rangle_\text{ideal}$ for the MQC method and $\langle\mathcal{C}\rangle = \left(1-2e\right)^{N}\langle\mathcal{C}\rangle_\text{ideal}$ for the parity oscillation method, where $N$ denotes the qubit number and $\langle\mathcal{C}\rangle_\text{ideal}\equiv \text{Tr}(\rho_\text{ideal}|0\rangle\langle 1|^{\otimes N})$ denotes the ideal correlation.
Therefore, the MQC technique has a clear advantage over the parity oscillation method in mitigating the impact of readout errors, which is verified in our experiment (Extended data Fig.~\ref{ex_fig:parity_vs_MQC}).
See Supplementary Sec.~II.C for a detailed analysis.

\vspace{.6cm}
\noindent\textbf{\large{}Data availability}\\
The data presented in the figures and that support the other findings of this study will be made publicly available upon its publication. 

\vspace{.5cm}
\noindent\textbf{Acknowledgement} 
We thank Antonio Ac\'in, Vedran Dunjko, Marc-Olivier Renou, Jens Eisert, Benjamin Schiffer, and Dong Yuan for helpful discussions. 
The device was fabricated at the Micro-Nano Fabrication Center of Zhejiang University. 
We acknowledge support from the Innovation Program for Quantum Science and Technology (Grant Nos.~2021ZD0300200 and 2021ZD0302203), the National Natural Science Foundation of China (Grant Nos.~12174342, 92365301, 12274367, 12322414, 12274368, 12075128, and T2225008), the National Key R\&D Program of China (Grant No.~2023YFB4502600), and the Zhejiang Provincial Natural Science Foundation of China (Grant Nos.~LDQ23A040001, LR24A040002). 
W.L., W.J., Z.L., Z.-Z.S., and D.-L.D. are supported in addition by Tsinghua University Dushi Program, and the Shanghai Qi Zhi Institute.
P.E. and J.T. acknowledge the support received by the Dutch National Growth Fund
(NGF), as part of the Quantum Delta NL programme. 
P.E. acknowledges the support received through the NWO-Quantum Technology
programme (Grant No.~NGF.1623.23.006).
J.T. acknowledges the support received from the European Union's Horizon Europe research and innovation programme through the ERC StG FINE-TEA-SQUAD (Grant No.~101040729). 
This publication is part of the `Quantum Inspire - the Dutch Quantum Computer in the Cloud' project (with project number [NWA.1292.19.194]) of the NWA research program `Research on Routes by Consortia (ORC)', which is funded by the Netherlands Organization for Scientific Research (NWO).
The views and opinions expressed here are solely
those of the authors and do not necessarily reflect those of the funding institutions. Neither
of the funding institutions can be held responsible for them.

\vspace{.3cm}
\noindent\textbf{Author contributions}  
K.W. and S.X. carried out the experiments under the supervision of C.S. and H.W.. J.C. and X.Zhu designed the device and H.L. fabricated the device, supervised by H.W.. W.L., M.H., P.E., J.T., and D.-L.D. conducted the theoretical analysis. All authors contributed to the analysis of data, the discussions of the results, and the writing of the manuscript.

\vspace{.3cm}
\noindent\textbf{Competing interests}  All authors declare no competing interests.

\setcounter{figure}{0}
\renewcommand{\theHfigure}{A.Abb.\arabic{figure}}
\renewcommand{\figurename}{Extended Data Fig.}

\vspace{.3cm}
\noindent\textbf{Extended data}

\begin{figure*}[htb]
\includegraphics[width=1\linewidth]{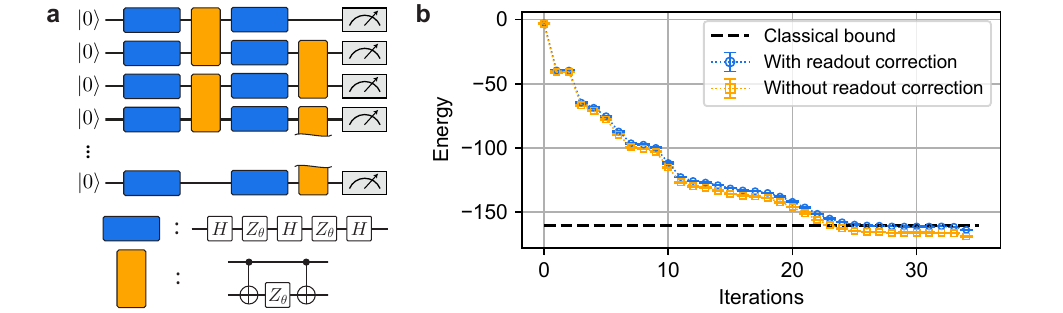}
\caption{\textbf{Observation of Bell correlations in a $\bold{1}$D chain.}
\textbf{a}, The variational quantum circuit for detecting the Bell correlations, where 
variational parameters are specified as $\theta$ in the virtual Z gate~\cite{PhysRevA.96.022330}. \textbf{b}, The optimization trajectories of the XXZ-type Hamiltonian with $\Delta=2$ and $\epsilon=0.95$. }
\label{ex_fig:XXZ}
\end{figure*}

\begin{figure*}[htb]
\includegraphics[width=1\linewidth]{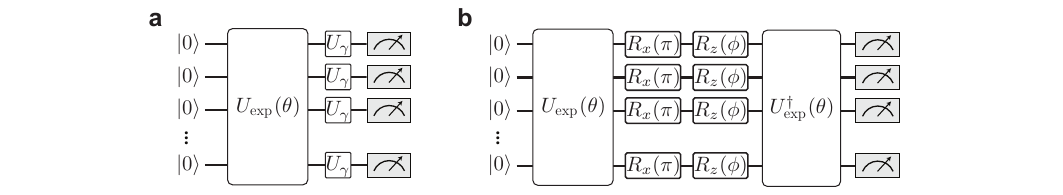}
\caption{\textbf{Quantum circuits for measuring multi-qubit correlations.}
\textbf{a}, Circuit for measuring parity oscillation. We add a single-qubit gate $U_\gamma=\frac{\sqrt{2}}{2} \left( {\begin{matrix} 1&e^{-i\gamma} \\ -e^{i\gamma} & 1\end{matrix}} \right)$ to each qubit at the end of the variational circuit $U_{\text{exp}}\left(\theta \right)$ depicted in Figure 3(a) of the main text, which
brings the axis defined by the operator $\mathcal{P}(\gamma)=\sin(\gamma) \sigma_y + \cos(\gamma) \sigma_x$ to the $z$ axis. Subsequently, we perform simultaneous measurements on all qubits and calculate the parity expectations with different $\gamma$ values. \textbf{b}, Circuit for measuring MQC. After applying the variational circuit $U_{\text{exp}}\left(\theta \right)$, we add a single-qubit $R_x(\pi)$ gate and $R_z(\phi)$ gate to each qubit followed by a reversal circuit $U^\dagger_{\text{exp}}\left(\theta \right)$. Then we measure the probability of the system being in the $\ket{00\dots0}$ state.}
\label{ex_fig:circuit}
\end{figure*}

\begin{figure*}[htb]
\includegraphics[width=1\linewidth]{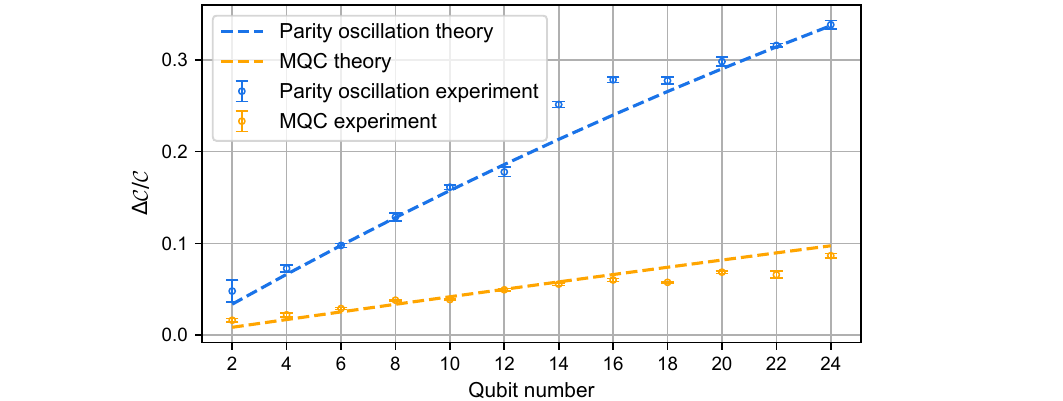}
\caption{\textbf{The effect of readout error.} 
We plot the effect of readout error with different qubit numbers estimated based on experimental data (dots with error bars) and theoretical calculation (dashed lines) for both the MQC (orange) and parity oscillation (blue) methods.
For the experimental data, the effect of readout error is estimated as the ratio between the correlation degradation due to the readout error ($\Delta \mathcal{C}$) and the correlation after mitigating the readout error ($\mathcal{C}$).
For the theoretical data, the effect of readout error can be estimated as $1-(1-e)^{N/2}$ and $1-(1-2e)^N$ for the MQC and parity oscillation methods, respectively.
We use an error rate of $0.0085$ for both the $\ket{0}$ and $\ket{1}$ states to estimate the theoretical values. 
}
\label{ex_fig:parity_vs_MQC}
\end{figure*}

\bibliography{QMLBib}

\end{document}


\captionsetup[figure]{name={Supplementary Figure}, singlelinecheck=off,,labelfont=bf,labelsep=period,justification=raggedright} 
\captionsetup[table]{name={Supplementary Table}, singlelinecheck=off,,labelfont=bf,labelsep=period,justification=raggedright} 

\setcounter{equation}{0}
\setcounter{figure}{0}
\setcounter{table}{0}

\renewcommand{\theequation}{S\arabic{equation}}
\renewcommand{\thefigure}{S\arabic{figure}}

\title{Supplementary: Probing many-body Bell correlation depth with superconducting qubits}

\maketitle
\tableofcontents

\section{Theoretical details}
\label{supp:theory}

\subsection{Bell nonlocality: An introduction}
\label{supp:nonlocality}

Bell inequalities provide a criterion to experimentally test whether the outputs from quantum mechanics can be explained by a local hidden variable theory. 
The violation of a Bell inequality signals nonlocality, which has been demonstrated in loophole-free Bell experiments across various platforms~\cite{Hensen2015Loopholefree,Giustina2015SignificantLoopholeFree,Shalm2015Strong,Li2018Test,Rosenfeld2017EventReady,Storz2023Loopholefree}. 
Furthermore, certifying nonlocality in an experimental setup serves as compelling evidence that its behavior is intrinsically quantum. 
In this subsection, we introduce the Bell inequalities used in this work, as well as the concepts of multipartite Bell scenarios and Bell nonlocality depth.

\subsubsection{Clauser-Horne-Shimony-Holt inequality}

\label{supp:bell}

The predictions of quantum mechanics are incompatible with the assumption of a local hidden variable theory. According to Bell's theorem~\cite{Einstein1935Can}, this distinction can be measured via Bell's inequalities~\cite{Bell1964Einstein}. One example is the Clauser-Horne-Shimony-Holt (CHSH) inequality~\cite{Clauser1969Proposed}.
For simplicity, let us consider the scenario involving two spatially separated parties, denoted as $A$ and $B$. 
Each party can perform one of two measurements, represented by $x \in \{0,1\}$ for party $A$ and $y \in \{0,1\}$ for party $B$. 
Each measurement has two outcomes, which are labeled by $a,b\in \{-1,+1\}$ for $A$ and $B$, respectively. 
This process is characterized by the conditional probability distribution $P(ab\vert xy)$ of obtaining outcomes $(a,b)$ when performing measurements $(x,y)$. 

Consider the expression 
\begin{equation}
    \mathcal{I} = \langle A_0B_0 \rangle + \langle A_0B_1 \rangle + \langle A_1B_0 \rangle - \langle A_1B_1 \rangle,
\end{equation}
where $\langle A_xB_y \rangle = \sum_{a,b} abP(ab \vert xy)$ is the expectation value of outcomes $(a, b)$ when choosing measurements $(x, y)$. 
If $P(ab \vert xy)$ agrees with a local hidden variable model, meaning there exists a probability distribution $p(\lambda)$ such that $P(ab \vert xy)= \sum_{\lambda}p(\lambda)P(a \vert x,\lambda)P(b \vert y,\lambda)$, then $\lvert \mathcal{I}\rvert \leq 2$, which is the well-known CHSH inequality. 
In this work, we choose the $\mathcal{I}\geq -2$ formulation, since we are focusing on a minimization task.
However, if $P(ab \vert xy)$ is not of local correlations, the CHSH inequality can be violated.
Specifically, let these two parties share the bipartite state 
\begin{equation}\ket{\psi} = \frac{1}{\sqrt{2}}(\ket{01} - \ket{10}).\nonumber
\end{equation}
Now party $A$ performs the measurements $A_0 = \sigma_z$, $A_1 = \sigma_x$, and party $B$ performs the measurements $B_0 = \frac{1}{\sqrt{2}}(\sigma_z + \sigma_x)$ and $B_1 = \frac{1}{\sqrt{2}}(\sigma_z - \sigma_x)$, where $\sigma_x, \sigma_z$ are Pauli operators. 
Then, we have $\langle A_xB_y \rangle = -\frac{1}{\sqrt{2}}(-1)^{x\cdot y}$, for $x, y \in \{0, 1\}$. Further, we can obtain $\mathcal{I} = -2\sqrt{2} < -2$, which violates the CHSH inequality. Thus, CHSH inequality provides a way to distinguish between the set of quantum correlations and local correlations under the assumption of local hidden variables. 

\subsubsection{Gisin's elegant inequality and its generalization}
\label{supp:gisin}

\begin{figure}
    \centering
    \includegraphics[width=8cm]{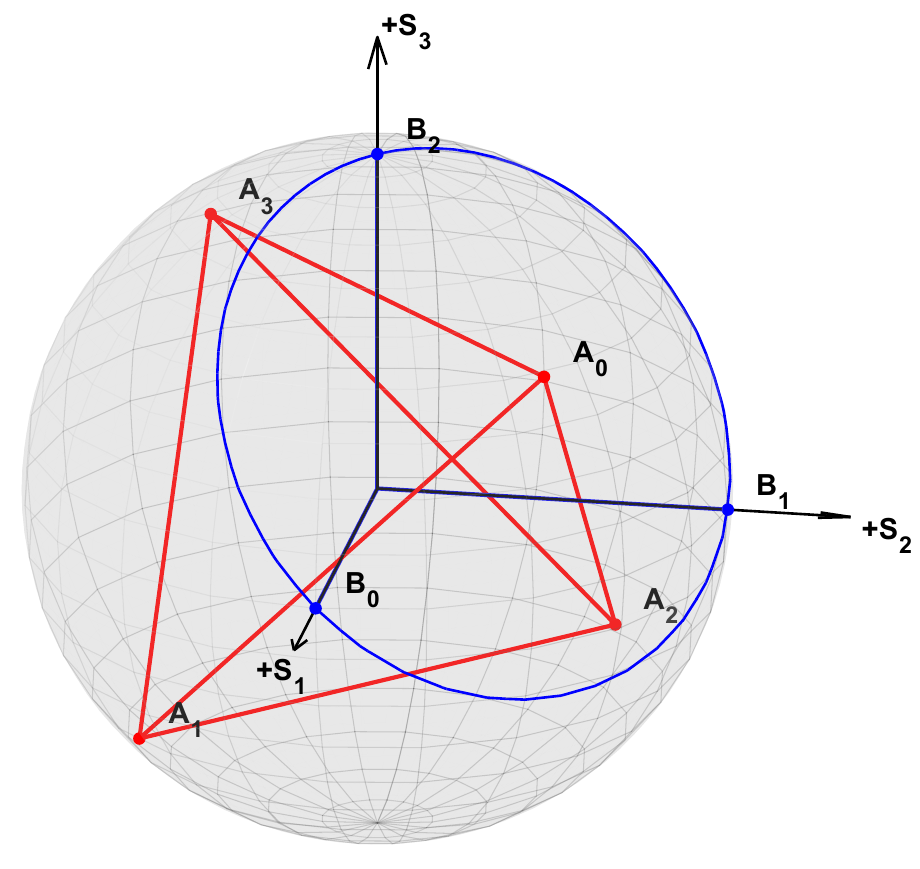}
    \caption{Measurement settings represented on the Poincar\'{e} sphere for Gisin's elegant inequality. 
    For party $A$, the four measurements are the vertices of the tetrahedron: $\mathrm{A_0}=\frac{1}{\sqrt{3}}(1,1,1)$, $\mathrm{A_1} = \frac{1}{\sqrt{3}}(1,-1,-1)$, $\mathrm{A_2} = \frac{1}{\sqrt{3}}(-1,1,-1)$, and $\mathrm{A_3} = \frac{1}{\sqrt{3}}(-1,-1,1)$ (points in red color). 
    For party $B$, the three measurements are mutually orthogonal: $\mathrm{B_0} = (1,0,0)$, $\mathrm{B_1} = (0,1,0)$, and $\mathrm{B_2} = (0,0,1)$ (points in blue color), which are parallel with the three axes.}
    \label{fig:gisin_measure}
\end{figure}

Gisin's elegant inequality involves two distant parties $A$ and $B$, with four measurements for party $A$ and three measurements for party $B$, respectively. 
All measurements produce two outcomes, $\pm 1$~\cite{Gisin2009Bell}. 
Let $ \langle A_xB_y \rangle=\sum_{a,b}abP(ab|xy)$ again be the expectation value of the product of the outcomes of $A$'s $x$th measurement and $B$'s $y$th measurement. 
Gisin's elegant inequality, denoted as $\mathcal{I}\geq \Bc$, reads as follows,
\begin{equation}
    \begin{split}
          \label{eq:gisin_ineq}
    \mathcal{I}=&\langle A_0B_0 \rangle+\langle A_1B_0 \rangle-\langle A_2B_0 \rangle-\langle A_3B_0 \rangle+\langle A_0B_1 \rangle-\langle A_1B_1 \rangle+\langle A_2B_1 \rangle-\langle A_3B_1 \rangle \\
    &+\langle A_0B_2 \rangle-\langle A_1B_2 \rangle-\langle A_2B_2 \rangle+\langle A_3B_2 \rangle \\
    &\geqslant \Bc = -6,
    \end{split}
\end{equation}
where $\Bc=-6$ is the classical bound. 
Note that the inequality $\mathcal{I} \geq -6$ is not a facet of the polytope of local correlations~\cite{Gisin2009Bell}. 
The maximal quantum violation is $-4\sqrt{3}\approx 6.93$, which is attained when party $A$ and $B$ share the state $\ket{\psi}=\frac{1}{\sqrt{2}}(\ket{01}-\ket{10})$~\cite{Acin2016Optimal}. 
In the case of the maximal violation, the four measurements of party $A$ are
\begin{equation}
\begin{aligned}
A_0&=\frac{\sigma_x+\sigma_y+\sigma_z}{\sqrt{3}},  \quad
A_1=\frac{\sigma_x-\sigma_y-\sigma_z}{\sqrt{3}}, \\
A_2&=\frac{-\sigma_x+\sigma_y-\sigma_z}{\sqrt{3}},  \quad
A_3=\frac{-\sigma_x-\sigma_y+\sigma_z}{\sqrt{3}}, \nonumber
\end{aligned}
\end{equation}
the three measurements of party $B$ are 
\begin{equation}
\begin{aligned}
B_0&=\sigma_x, \quad
B_1=\sigma_y,  \quad
B_2=\sigma_z,\nonumber
\end{aligned}
\end{equation}
where $\sigma_x,\sigma_y,\sigma_z$ are Pauli operators. We can represent this measurement setting on the Poincar\'{e} sphere as shown in Fig.~\ref{fig:gisin_measure}. 
The four measurements of party $A$ are the vertices of a tetrahedron and the three measurements of party $B$ are mutually orthogonal. 
The Bell operator of $\mathcal{I}$ in Eq.~\eqref{eq:gisin_ineq} is 
\begin{align}
    \label{eq:Gisin_bell_op}
    \mathcal{B}=\frac{4}{\sqrt{3}}(\sigma_x^{A}\sigma_x^{B}+\sigma_y^{A} \sigma_y^{B}+\sigma_z^{A}\sigma_z^{B}),
\end{align}
and one can check that $\bra{\psi}\mathcal{B}\ket{\psi}=-4\sqrt{3}$ is the maximal quantum violation.

With the same measurement setting, we use a modified version of Gisin's elegant inequality
\begin{equation}
\label{eq:XXZ_delta}
\begin{aligned}
   \mathcal{I}_{\Delta}&=\langle A_0 B_0 \rangle + \langle A_1 B_0 \rangle - \langle A_2 B_0 \rangle - \langle A_3 B_0 \rangle+\langle A_0 B_1 \rangle - \langle A_1 B_1 \rangle + \langle A_2 B_1 \rangle - \langle A_3 B_1 \rangle
   \\&+\Delta(\langle A_0 B_2 \rangle - \langle A_1 B_2 \rangle - \langle A_2 B_2 \rangle + \langle A_3 B_2 \rangle)
   \\&\geqslant -(2|\Delta|+|\Delta+2|+|\Delta-2|),\nonumber
\end{aligned}
\end{equation}
as introduced in the main text, where $\Delta$ is a real number~\cite{Tura2017Energy}. Now its corresponding Bell operator becomes 
\begin{align}
    \label{eq:Gisin_bell_op_general}
    \mathcal{B}_{\Delta}=\frac{4}{\sqrt{3}}(\sigma_x^{A}\sigma_x^{B}+\sigma_y^{A} \sigma_y^{B}+\Delta\sigma_z^{A}\sigma_z^{B}).
\end{align} 
The ground state energy of this operator, i.e., the minimal eigenvalue, is
$\frac{4}{\sqrt{3}}\Delta$ for $\Delta \leq -1$ and $-\frac{4}{\sqrt{3}}(2+\Delta)$ for $\Delta>-1$.

\subsubsection{Multipartite scenario}
\label{supp:multi}

To explore the Bell nonlocality and entanglement properties in a larger and more complex system, it is essential to study multipartite Bell inequalities.
Consider a Bell scenario where $N$ spatially separated parties $A_1,A_2, \dots,A_N$ share some $N$-partite resource. 
On their share of this resource, each party $A_i$ can perform one of $m$ measurements and each measurement can yield $d$ possible outcomes. 
We refer to this scenario as a Bell scenario $(N,m,d)$. 
For party $A_i$, we label the measurements choices $x_i \in [m]=\{0,\dots,m-1\}$  and its outcomes $a_i \in [d]= \{0,\dots,d-1\}$, respectively. The correlations between results are described by a collection of conditional probability distributions
\begin{align}
    \label{eq:correlations}
    \{P(a_1,\dots,a_{N}|x_1,\dots,x_{N
    })\}_{a_1,\dots,a_{N};x_1,\dots,x_{N}},  
\end{align}
where $P(a_1,\dots,a_{N}|x_1,\dots,x_{N
})=:P(\mathbf{a}\vert \mathbf{x})$ is the probability of obtaining outcomes $\mathbf{a}:=a_1,\dots,a_{N}$ by performing measurements $\mathbf{x}:=x_1,\dots,x_{N
}$. 
We can order these probabilities in \eqref{eq:correlations} into a vector $\vec{p}$ with $(md)^N$ components, that is,  
\begin{align}
    \vec{p}=	\{P(\mathbf{a}\vert \mathbf{x})\}_{\mathbf{a};\mathbf{x}} \in \mathbb{R}^{(md)^N} . 
\end{align}

In characterizing multipartite correlations, one typically studies whether they are compatible with a given physical model.
Three notable examples are the no-signaling principle, local hidden variable models, and quantum theory.

\textbf{Non-signalling correlations} are defined by the no-signaling principle which states that all parties are spatially separated and cannot communicate instantaneously. 
This ensures that the marginals observed by any subset of parties remain consistent and well-defined, i.e. any subset of parties does not depend on the measurement choices of the other parties~\cite{cirelson_quantum_1980,popescu_quantum_1994}.
In terms of probability distributions in \eqref{eq:correlations}, the no-signaling principle can be expressed as
\begin{align}
    \label{eq:const_NS}
    \sum_{a_i}	P(a_1,\dots,a_i,\dots,a_{N}|x_1,\dots,x_i,\dots,x_{N})=\sum_{a_i}P(a_1,\dots,a_i,\dots,a_{N}|x_1,\dots,x'_i,\dots,x_{N})
\end{align}
for all $x_i \neq x'_i$,$a_1,\dots,a_{i-1},a_{i+1},\dots,a_{N}$ and $x_1,\dots,x_{i-1},x_{i+1},\dots,x_{N}$.
Thus, $P(a_{i_1},\dots,a_{i_l}\vert x_{i_1},\dots,x_{i_l})$ is well defined on any subset $\{i_1,\dots,i_l\}\subseteq \{1,\dots,N\}$. 
Moreover, the probability distributions in \eqref{eq:correlations} need to satisfy the $m^N$ affine-linear equations enforcing the normalization of the probabilities
\begin{equation}\label{eq:probsSumToOne}
    \sum_{a_1,\dots,a_N}P(a_1,\dots,a_N | x_1,\dots,x_N) = 1, \forall x_1,\dots,x_N,
\end{equation}
as well as the inequalities $P(\mathbf{a}\vert \mathbf{x})\geq 0$. 
The feasibility region for $\vec{p}$ satisfying the no-signaling principle is a polytope, which we denote by $\mathcal{NS}$.

The second class of correlations is defined by \textbf{local hidden variable models} (LHVMs).
To detect Bell nonlocality, we need to find Bell inequalities separating the polytope $\mathcal{L}$ of local correlations described by LHVMs from the convex set of quantum correlations $\mathcal{Q}$.
The set of probability distributions in Eq.~\eqref{eq:correlations} that admit the LHVM is formed by those $\vec{p}$ in which all $P(\mathbf{a}\vert \mathbf{x})$ can be written as 
\begin{align}
        \label{eq:p_local_corr}
    P(a_1,\dots,a_{N}|x_1,\dots,x_{N
    })=\sum_{\lambda}p(\lambda)\prod_{i=1}^{N}P(a_i | x_i,\lambda),
\end{align}
where $\lambda$ is some hidden variable distributed according to a probability distribution $p(\lambda)$. 
The compatibility region for probability distributions that admit an LHVM description forms a polytope which we denote by $\mathcal{L}$. 
The vertices of $\mathcal{L}$ factorize as a product of deterministic correlations as follows:
\begin{align}
    \label{eq:LHV}
    P(a_1,\dots,a_{N}|x_1,\dots,x_{N
    })=\prod_{i=1}^{N}P(a_i|x_i),
\end{align}
where each $P(a_i|x_i)$ is deterministic, i.e., $P(a_i|x_i)=\delta(a_i-\alpha_{i,x_i})$, $\delta$ is the Kronecker delta function and $\alpha_{i,x_i} \in [d]$~\cite{FinePRL1982}. 

Finally, we consider the \textbf{set of quantum correlations} $\mathcal{Q}$, where each element $P(\mathbf{a}\vert \mathbf{x})$ can be represented by Born's rule:
\begin{align}
    \label{eq:quantumcorrelations}
    P(a_1,\dots,a_{N}\vert x_1,\dots,x_{N})=\Tr[\rho_N(\mathcal{M}^{a_1}_{1,x_1}\otimes \dots \otimes \mathcal{M}^{a_{N}}_{N,x_{N}})],
\end{align}
where $\rho_N$ is an $N$-partite quantum state and $\mathcal{M}_{i,x_i}^{a_i}\succeq 0$ is POVM element corresponding to the $x_i$-th measurement with outcome $a_i$ performed by party $A_i$, satisfying the normalization condition $\sum_{a_i}\mathcal{M}_{i,x_i}^{a_i}=\mathbb{I}$. 
Note that $\mathcal{Q}$ is a convex set but not a polytope~\cite{PhysRevA.97.022104}. 
Moreover, we know $\mathcal{L} \subsetneq \mathcal{Q} \subsetneq \mathcal{NS}$~\cite{Brunner2014Bell}. 

For convenience, we can represent the correlations in Eq.~\eqref{eq:correlations} using expectation values instead of conditional probabilities. 
We restrict the number of outcomes to $d=2$, which is the case considered in this work. 
First we write the probabilities $P(a_1,\dots,a_{N}|x_1
,\dots,x_{N})$ using $N$-dimensional discrete Fourier transformation as follows,
\begin{align}
    \label{eq:correlators}
    \langle A_{1,x_1}^{(k_1)}\dots A_{N,x_N}^{(k_N)}\rangle=\sum_{(a_1,\dots,a_N
        )\in\{0,1\}^N}(-1)^{(k_1,\dots,k_N)\cdot (a_1,\dots,a_N)}P(a_1,\dots,a_N|x_1,\dots,x_N),
\end{align}
where $k_1,\dots,k_N\in [2]$. Note that by using the no-signaling principle, one can obtain that when $k_i=0$,
\begin{align}
    \langle A_{1,x_1}^{(k_1)}\dots A_{N,x_N}^{(k_N)}\rangle=\langle A_{1,x_1}^{(k_1)}\dots A_{{i-1},x_{i-1}}^{(k_{i-1})} A_{{i+1},x_{i+1}}^{(k_{i+1})}\dots  A_{N,x_N}^{(k_N)}\rangle,
\end{align}
and $\langle  A_{1,x_1}^{(0)}\dots A_{N,x_N}^{(0)}\rangle=1$ for any $(x_1,\dots,x_N)$.

Finally, for the Bell scenario $(N,m,d)$, the general form of a Bell inequality $\mathcal{I}_{N,m,d}-\Bc \geq 0$ is 
\begin{align}
    \label{eq:bellineq}
    &&\mathcal{I}_{N,m,d}:=\sum_{x_1,\dots,x_N=0}^{m-1}\sum_{k_1,\dots,k_N=0}^{1} \alpha_{x_1,\dots,x_N}^{(k_1,\dots,k_N)} \langle  A_{1,x_1}^{(k_1)}\dots A_{N,x_N}^{(k_N)}\rangle,
\end{align}
where the correlators $\langle  A_{1,x_1}^{(k_1)}\dots A_{N,x_N}^{(k_N)}\rangle$ are defined in Eq.~\eqref{eq:correlators}, and $\Bc$ is the classical bound. 
Such Bell inequalities constrain the local polytope $\mathcal{L}$. 
The violation of a Bell inequality $\mathcal{I}_{N,m,d}  -\Bc \geq 0$ with $\Bc \coloneq \min_{P(\mathbf{a}\vert \mathbf{x}) \in \mathcal{L}} \mathcal{I}_{N,m,2}$ detects Bell nonlocality. 
The inequality $\mathcal{I}_{N,m,2} -\Bq\geq  0$,  where $\Bq \coloneq \inf_{P(\mathbf{a}\vert \mathbf{x}) \in \mathcal{Q}} \mathcal{I}_{N,m,2}$ is called the Tsirelson bound~\cite{cirelson_quantum_1980}.

\subsubsection{Bell nonlocality depth}
\label{supp:depth}

The violation of Bell inequalities $\mathcal{I}_{N,m,d} \geq \Bc$ signals nonlocality, implying that the multipartite probability distribution $P(\mathbf{a}\vert \mathbf{x})$ cannot be written in the form of Eq.~\eqref{eq:p_local_corr}. 
However, the violation of a Bell inequality does not directly provide any information about the structure of the correlations. 
For example, a single singlet in a product state would be enough to violate a Bell inequality~\cite{gisin1991bell}. 
More concretely, consider the tripartite probability distribution $P(a_1a_2a_3\vert x_1x_2x_3)$. 
If, for example, it can be written as a product of probability distributions such as $P(a_1a_2\vert x_1x_2)P(a_3\vert x_3)$, and $P(a_1a_2\vert x_1x_2)$ describes nonlocal correlations, the whole distribution is considered nonlocal. 
Thus, a multipartite probability distribution can manifest in various classes of nonlocality. 
The concept of genuine multipartite nonlocality helps us to classify them~\cite{Bancal2009Quantifying,Baccari2019Bell,Bernards2023Bell}. 

To explain nonlocality depth, we first formulate the concept of $k$-producible probability distributions.
For $N$ parties sharing some multipartite correlations, we are interested in quantifying the number of parties that share genuine multipartite nonlocality. 
To be more precise, consider these $N$ parties are in a partition $L_k$ of the set of indices $S=\{1,\dots,N\}$ into $L$ pairwise disjoint nonempty subset $S_i$, where $S_i \cap S_j =\varnothing$ if $i \neq j$ and the cardinality of each subset $|S_i| \leq k$. 
Then we say an $N$-partite probability distribution is $k$-producible with respect to the partition  $L_k$ if and only if it has the following decomposition: 
\begin{align}
    \label{eq:k-producible_prob}
    P_{L_k}(\mathbf{a}\vert \mathbf{x})=\sum_{\lambda}p(\lambda) \prod_{i=1}^LP(\mathbf{a}_{S_i}\vert \mathbf{x}_{S_i},\lambda),
\end{align}
where $\mathbf{a}_{S_i}, \mathbf{x}_{S_i}$ are the tuples of outcomes and measurement choices corresponding to the parties in $S_i$. 
The largest number of parties exhibiting nonlocal correlations is the largest cardinality of the subset $S_i$ which is at most $k$. 
Notice that we restrict the distributions $P(\mathbf{a}_{S_i}\vert \mathbf{x}_{S_i},\lambda)$ to satisfy the non-signaling principle in Eq.~\eqref{eq:const_NS}.
However, there are many ways to partition the set of indices $S$ as $L_k$, we can generalize the definition of $k$-producible by using a convex combination of Eq.~\eqref{eq:k-producible_prob}. 
To list all the possible partitions $L_k$ of $S$ for fixed $k$, note that the number of subsets $L$ can be varied, we define $\mathcal{P}_k$ to be the set of all possible partitions $L_k$. 
Now, $P(\mathbf{a} \vert \mathbf{x})$ is $k$-producible if and only if
\begin{align}
    \label{eq:k-producible_prob_general}
     P(\mathbf{a}\vert \mathbf{x})=\sum_{L_k \in \mathcal{P}_k} q_{L_k}P_{L_k}(\mathbf{a}\vert \mathbf{x}),
\end{align}
where $P_{L_k}(\mathbf{a} \vert \mathbf{x})$ are the correlations as defined in \eqref{eq:k-producible_prob} corresponding to the partition $L_k$. 
The minimum $k$ of $P(\mathbf{a}\vert \mathbf{x})$ which can be written in the form of Eq.~\eqref{eq:k-producible_prob_general} is the so-called nonlocality depth~\cite{Bancal2009Quantifying,Baccari2019Bell,Bernards2023Bell}. This is the main concept we focus on in this work.

The set of all the $k$-producible probability distribution is a polytope that we denote $\mathcal{L}_k$. 
To characterize this polytope $\mathcal{L}_k$, we need to consider all the possible partitions $L_k$ and their corresponding product of probability distributions $P(\mathbf{a}\vert \mathbf{x})$. 
Similarly to the local polytope $\mathcal{L}$, the vertices of the polytope $\mathcal{L}_k$ are as follows:
\begin{align}
    \label{eq:L_k_vertices}
    P(\mathbf{a}\vert \mathbf{x})=\prod_{i=1}^L P(\mathbf{a}_{S_i}\vert \mathbf{x}_{S_i}),
\end{align}
where $P(\mathbf{a}_{S_i}\vert \mathbf{x}_{S_i})$ is a vertex of non-signalling polytope with respect to $S_i$. 
Furthermore, the probability distributions $P(\mathbf{a}_{S_i}\vert \mathbf{x}_{S_i})$ can also be expressed in terms of correlators as illustrated in Eq.~\eqref{eq:correlators}.
The facets of the polytope $\mathcal{L}_k$ define tight Bell-like inequalities~\cite{cirelson_quantum_1980,Svetlichny1987Distinguishing,bancal2012framework}. 
Moreover, the violation of such Bell-like inequalities indicates a minimum Bell nonlocality depth of $k+1$, implying that the probability distribution is not $k$-producible. 
Thus, we can use these Bell-like inequalities to detect and quantify the nonlocality depth. 
However, characterizing the polytope $\mathcal{L}_k$ is an intractable task because of the need to list all the extreme points of $\mathcal{L}_k$ and the number of extreme points grows exponentially as a function of the number of parties $N$~\cite{fritz2012polyhedral,barrett2005nonlocal}.

We can still focus on certain inequalities constraining $\mathcal{L}_k$, for example, Mermin and Svetlichny inequalities~\cite{Mermin1990Extreme, Svetlichny1987Distinguishing}, to study nonlocality depth in the multipartite scenario. 
Although the Mermin and Svetlichny inequalities were derived in different contexts, they exhibit a form of equivalence when applied to detecting nonlocality depth~\cite{Mermin1990Extreme, Svetlichny1987Distinguishing, Barreiro2013Demonstration, Baccari2019Bell, Bancal2009Quantifying}. 
In the $(N,2,2)$ scenario, each party can perform two measurements, and each measurement $x_i$ yields two outcomes $a_i$, i.e., $x_i \in \{0,1\}$ and $a_i \in \{-1,+1\},i=1, \dots,N$. 
The Svetlichny expression in this case can be written as
\begin{align}
    \label{eq:Sve_ineq_N22}
    \mathcal{I}_{N}^{\text{Sv}}=2^{-\frac{N}{2}} \left[ \sum_{\mathbf{x} \vert ({s} = 0)}(-1)^{\frac{{s}}{2}} \langle \mathbf{A}\rangle_{\mathbf{x}}+ \sum_{\mathbf{x} \vert ({s} = 1)}(-1)^{\frac{{s-1}}{2}} \langle \mathbf{A}\rangle_{\mathbf{x}}\right],
\end{align}
where $\langle \mathbf{A}\rangle_{\mathbf{x}}=\langle A_{1,x_1}\dots A_{N,x_N}\rangle$ is the $N$-partite correlator, and $s$ is the parity of $\sum_{i}x_i$.
Similarly, the Mermin Bell expression can be reformulated as 
\begin{align}
    \label{eq:Mer_ineq_N22}
     \mathcal{I}_{N}^{\text{Me}}=2^{-\frac{N-1}{2}}\left[ \sum_{\mathbf{x} \vert ({s}= 0)} (-1)^{\frac{s}{2}}\langle \mathbf{A}\rangle_{\mathbf{x}}\right].
\end{align}

Next, we can construct Bell operators corresponding to these two Bell expressions. 
The Svetlichny inequality is maximally violated when the $N$ parties share the state $\ket{\text{GHZ}}_N=\frac{1}{\sqrt{N}}\left(\ket{0}^{\otimes N}+\ket{1}^{\otimes N}\right)$ and the two measurements for each party $i\in \{1,\dots,N\}$ are 
\begin{eqnarray}
    A_{i,0}=\cos\left(-\frac{\pi}{4N}\right)\sigma_x+\sin\left(-\frac{\pi}{4N}\right)\sigma_y, \quad
    A_{i,1}=\cos\left(\frac{(2N-1)\pi}{4N}\right)\sigma_x+\sin\left(\frac{(2N-1)\pi}{4N}\right)\sigma_y.\nonumber
\end{eqnarray}
The Mermin inequality is also maximally violated by the state $\ket{\text{GHZ}}_N$ and the two measurements of each party $i\in\{1,\dots,N\}$ are $A_{i,0}=\sigma_x$ and $A_{i,1}=\sigma_y$. 
These two inequalities can detect genuine multipartite nonlocality effectively for any number of parties $N$, as the $(N-1)$-nonlocal bound is always larger than the maximal quantum bound~\cite{Baccari2019Bell}. 
 
By assigning the above measurement settings, the Bell operators of the Mermin and Svetlichny Bell inequalities are both expressed as
\begin{align}
    \label{eq:bell_operator_N22_Sve+Mer}
    \mathcal{B}^{\text{Me}}_{N}=\mathcal{B}^{\text{Sv}}_{N}=\mathcal{B}_{N}=2^{\frac{N-1}{2}}[(|0\rangle\langle 1|)^{\otimes N}+(|1\rangle\langle0|)^{\otimes N}].
\end{align}
The criterion for detecting Bell correlation depth follows from Ref.~\cite{Baccari2019Bell}:
\begin{align}
    \label{eq:bell_op_bound}
    \langle \mathcal{B}_{N} \rangle \geq -2^{(N-\lceil N/k \rceil)/2},
\end{align}
where $k$ is the Bell correlation depth. 
The violation of the inequality in Eq.~\eqref{eq:bell_op_bound} certifies that the Bell correlation depth is at least $k+1$. 

\subsection{Variational quantum circuits}
\label{supp:vqc}

With the rapid development in the field of quantum computation, various quantum algorithms have been developed to tackle practical problems in various areas including combinatorial optimization, cryptography, and finance~\cite{Cerezo2021Variational,Bharti2022Noisy}. 
While quantum algorithms may bring potential advantages to these areas, many of them demand quantum hardware capabilities far beyond noisy intermediate-scale quantum (NISQ) devices. 
To alleviate this requirement, the hybrid quantum-classical framework has emerged as one of the leading strategies for NISQ devices. 
Within this framework, variational quantum algorithms (VQAs) have drawn broad interest over recent years~\cite{Peruzzo2014Variational,Tilly2022Variational,Cerezo2021Variational,Bharti2022Noisy}. 
In this subsection, we introduce the basic framework and optimization strategies of our VQAs.

\subsubsection{Basic framework}
\label{supp:vqa_framework}

Our VQA aims to approximate the ground state energy of a quantum system Hamiltonian via a variational quantum circuit (VQC), which shares the idea of variational quantum eigensolvers~\cite{Peruzzo2014Variational}. 
For any given quantum state $|\psi\rangle$, the expectation value of Hamiltonian $H$ is lower bounded by the ground state energy $E_0$, i.e.,
\begin{equation}
\langle\psi|\mathcal{H}| \psi\rangle \geq E_0 .
\end{equation}
Then, the ground state energy can be approximately obtained by minimizing the expectation value of $\mathcal{H}$ with respect to the ansatz wavefunction $|\psi(\bm{\theta})\rangle$ parameterized by $\bm{\theta}$:
\begin{equation}
E(\bm{\theta}) = \langle\psi(\bm{\theta})|\mathcal{H}|\psi(\bm{\theta})\rangle .
\end{equation}
In the VQA scheme, the $N$-qubit quantum state $|\psi\rangle$ is prepared by applying a parameterized quantum unitary $U(\bm{\theta})$ to a initial state $|\bm{0}\rangle = |0\rangle^{\otimes N}$, which can be expressed as:
\begin{equation}
|\psi(\bm{\theta})\rangle = U(\bm{\theta}) |\bm{0}\rangle .
\end{equation}
In experiments, the quantum unitary $U(\bm{\theta})$ can be realized by a quantum circuit with the variational gate parameter $\bm{\theta}$. 
During the VQA procedure, $\bm{\theta}$ is optimized to minimize $E(\bm{\theta})$.
\begin{figure}[t]
\includegraphics[width=0.8\textwidth]{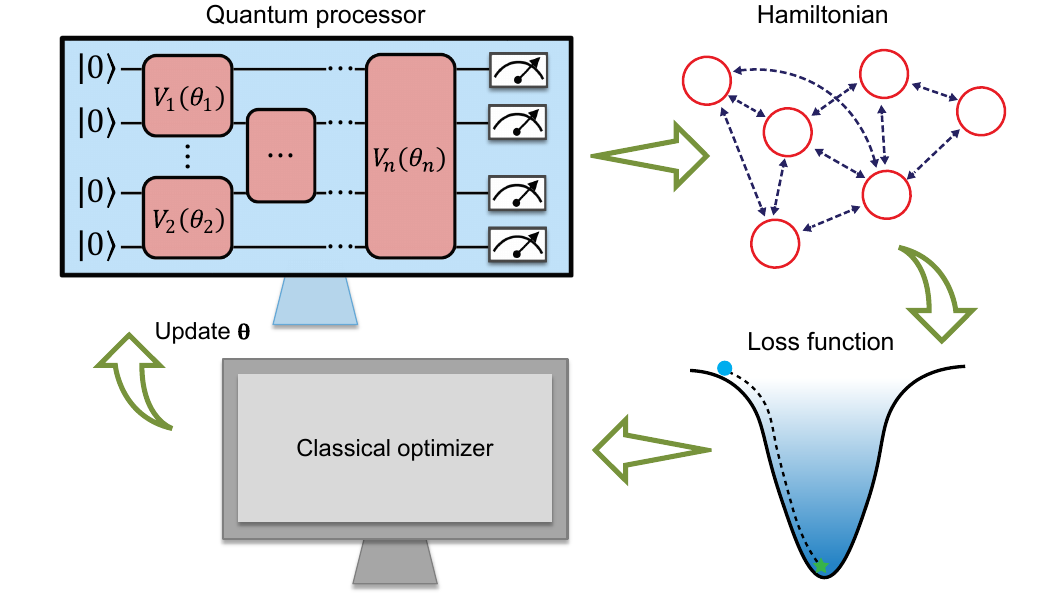}
\caption{
\textbf{Framework of the variational quantum algorithms.} 
In VQC, quantum states are prepared on a quantum processor with parameterized gates. The expectation value of $H$ is acquired through quantum measurements, followed by classical post-calculation. The obtained result is then fed into a classical optimizer, which iteratively determines new parameter values for the quantum processor.
}
\label{fig:vqa}
\end{figure}

\subsubsection{Optimization strategies}
\label{supp:opt}
The aforementioned variational ansatz is optimized in experimental quantum devices. 
To this end, we first choose an appropriate loss function according to the practical problem and then minimize it to solve our problem. 
For variational quantum algorithms, our aim is to find the quantum state minimizing the expectation value of the target Hamiltonian, and the loss function is chosen as the expectation value of the energy
\begin{equation}
    L(\bm{\theta})=E(\bm{\theta})=\langle\psi(\bm{\theta})|\mathcal{H}|\psi(\bm{\theta})\rangle,
\end{equation}
where $\ket{\psi(\bm{\theta})}$ is the output state of the variational ansatz with parameters $\bm{\theta}$, and $\mathcal{H}$ is the target Hamiltonian. 
If we fix {{the initial state as $\ket{\boldsymbol{0}}$}}, and represent the variational ansatz as a parametrized unitary $U(\bm{\theta})$, the loss function becomes
\begin{equation}
    L(\bm{\theta})=E(\bm{\theta})=\langle\bm{0}|U^\dagger(\bm{\theta})\mathcal{H}U(\bm{\theta})|\bm{0}\rangle.
\end{equation}
Then, the problem of finding the ground state for $\mathcal{H}$ is transformed to find the variational parameters $\bm{\theta}$ minimizing the loss function.

In practice, analytically finding the optimal variational parameters is very challenging. 
We usually update those parameters iteratively according to the instantaneous gradients until convergence, and the final parameters are regarded as the optimal solution.
In our work, the variational ansatz consists of single-qubit rotation gates and two-qubit controlled-Z/controlled-NOT gates.
The single qubit rotation gate has form $R(\theta_k)=e^{-i\theta_kP/2}$ with $P$ chosen from the single Pauli operators $\{\sigma_x,\sigma_y,\sigma_z\}$, and $\theta_k$ being the rotation angle. 
All variational parameters are encoded in the rotation angles. 
The corresponding derivative can be computed using the parameter shift rule~\cite{Mitarai2018Quantum,Schuld2019Evaluating}
\begin{equation}
    \frac{\partial L(\bm{\theta})}{\partial\theta_k}=\frac{L(\bm{\theta}\backslash\theta_k,\theta_k+\pi/2)-L(\bm{\theta}\backslash\theta_k,\theta_k-\pi/2)}{2},
\end{equation}
where $L(\bm{\theta}\backslash\theta_k,\theta_k\pm\pi/2)$ denotes the loss function where  $\theta_k$ is replaced by $\theta_k\pm\pi/2$ and the other parameters are unchanged. 
We remark that this analytical (exact and no bias) method can be implemented in current quantum devices to provide the gradients through measurements.
More optimization methods may be explored in the future~\cite{Wierichs2022General,Meyer2021Variational,Kyriienko2021Generalized,Banchi2021Measuring,Mari2021Estimating,Stokes2020Quantum,He2024TrainingFree}.

With gradients at hand, we can update the variational parameters using the gradient descent method.
As an illustrative example,
the direct gradient descent works by
\begin{equation}
    \bm{\theta}_{n+1}=\bm{\theta}_{n}-\epsilon\nabla_{\bm{\theta}}L(\bm{\theta}_{n}),
\end{equation}
where $\bm{\theta}_{i}$ is the parameters at $i$-th step and $\epsilon$ is a small scalar representing the update step length. 
In this work, we take the Adam[0.9, 0.999] optimizer to handle the obtained gradients, where the momentum-based method further improves the training performance~\cite{Kingma2014Adam}.

\subsubsection{Nonlocal correlation detection}
\label{supp:vqa4nonlocality}
Here, we introduce the basic ideas of utilizing variational quantum algorithms to detect Bell correlations. A typical Bell inequality has form
\begin{equation}
    \mathcal{I}-\Bc \geq 0,
\end{equation}
where $\mathcal{I}$ includes a collection of expectation values of correlators and $\Bc$ characterizes the limits of the local hidden variable model. 
For states admitting local-hidden-variable descriptions, the inequality is always valid. 
However, certain quantum states and measurements may yield correlations beyond the local-hidden-variable description and a Bell inequality would be violated. 
For a given set of measurements, the Bell inequality can be interpreted as a Bell operator and connected to a Hamiltonian~\cite{Tura2017Energy}.
If there exist quantum states violating such a Bell inequality, i.e., $\langle\mathcal{H}\rangle<\Bc$, the energy can be regarded as a witness of Bell nonlocality~\cite{Tura2017Energy}.

In this work, we extend this idea and adapt the variational quantum circuit as the variational ansatz to prepare low-energy states for the target Hamiltonian. 
We consider a standard Bell inequality in which the quantum correlators are expectation values for Pauli strings acting on different parties
\begin{equation}
    \mathcal{I}=\sum_{k}\langle \alpha_kM_k\rangle\geq \Bc,
\end{equation}
where $M_k$ is a Pauli string and $\alpha_k$ is the corresponding coefficient. The Hamiltonian is 
\begin{equation}
    H=\sum_k\alpha_kM_k.
\end{equation}

Within this framework, our general recipe for detecting the Bell correlations through variational quantum algorithms is as follows:
\begin{enumerate}
    \item We begin with a fixed initial state and a fixed quantum circuit ansatz including initialized variational parameters. At the initial stage, the output state typically does not violate the given inequality;
    \item We then use the optimization strategies described before to iteratively update the variational parameters, such that the energy of the output state is minimized.
\end{enumerate}
If the final minimal energy is smaller than $\Bc$, the corresponding Bell inequality is successfully violated and nonlocality is detected, while the maximal violation of the Bell inequality (the ground state) may not always be achieved due to experimental noises and limitations of the VQC's expressive power~\cite{Ren2022Experimental,Pan2023Deep,Li2022Quantum,Yu2023Expressibilityinduced}.

\section{Experimental details \label{sec:exp}}

\subsection{Device information}

\begin{figure}
    \centering
    \includegraphics[width=1\linewidth]{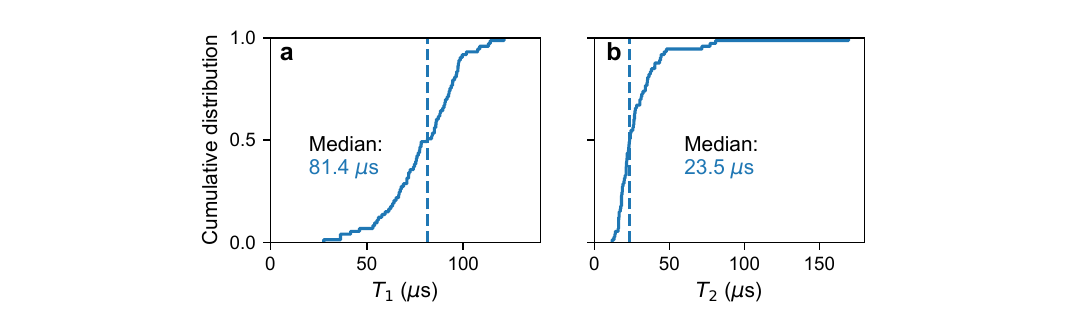}
    \caption{\textbf{Typical $T_1$ and $T_2$.} \textbf{a}, Cumulative distribution of $T_1$ across the $73$ qubits used on the superconducting quantum processor with a dashed line indicating a
    median value of \SI{81.4}{\micro\second}.
    \textbf{b}, Cumulative distribution of $T_2$ across the $73$ qubits used on the superconducting quantum processor with a dashed line indicating a
    median value of \SI{23.5}{\micro\second}.}
    \label{fig:T1T2}
\end{figure}

\begin{figure}
    \centering
    \includegraphics[width=1\linewidth]{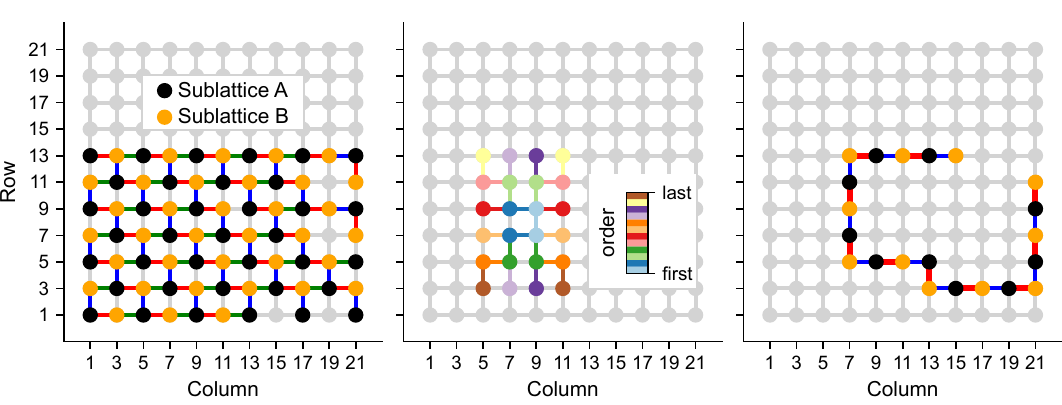}
    \caption{\textbf{Qubit layouts.} \textbf{Left}: The $73$ qubits used in the two-dimensional CHSH honeycomb model with qubit coordinates indicated as (row, column). \textbf{Middle}: The $24$ qubits used to detect Bell correlation depth. The order of training is indicated using colors, starting from $2$-qubit training and adding $2$ qubits each time. \textbf{Right}: The $21$ qubits used in the $1$D XXZ model. The red thick bars indicate the qubit pairs with strong coupling strength, while the blue thin bars indicate weak coupling strength.}
    \label{fig:qubit_topo}
\end{figure}

The superconducting quantum processor used in this work has $11 \times 11$ transmon qubits arranged in a square lattice and $220$ tunable couplers each inserted in-between neighboring two qubits~\cite{Xu2023nonAbelian}. 
Each qubit has individual microwave control and flux biases for single-qubit gates and frequency tunability, respectively. 
The coupling strength between any neighboring pair of qubits is dynamically tunable, allowing for high-fidelity two-qubit gates. 
Each qubit is dispersively coupled to a readout resonator for its state measurement. 
In our experiments, we use up to 73 qubits to construct the variational quantum circuit. 
The typical energy relaxation time $T_1$ and Hahn echo dephasing time $T_2$ of these qubits are shown in Fig.~\ref{fig:T1T2}, with median values of $T_1=\SI{81.4}{\micro\second}$ and $T_2=\SI{23.5}{\micro\second}$, respectively.
The qubit layouts used for the three different experiments in this work are shown in Fig.~\ref{fig:qubit_topo}.

\subsection{Experiment circuit}

The experimental circuit for the variational quantum algorithm contains multiple layers of single-qubit gates and two-qubit CNOT gates with different patterns. 
In the experiment, we realize arbitrary single-qubit gates by combining XY rotations and Z rotations, with the XY rotations implemented with $30$-ns long microwave pulses with DRAG pulse shaping~\cite{PhysRevLett.119.180511} at the qubits' respective idle frequencies, and the Z rotations realized by the virtual-Z gate scheme~\cite{PhysRevA.96.022330}. 
The CNOT gate is decomposed to a generic two-qubit controlled $\pi$-phase (CZ) gate sandwiched between two Hadamard gates. 
We realize the CZ gates by bringing $|11\rangle$ and $|02\rangle$ (or $|20\rangle$) of the qubit pairs in near resonance and tuning the coupler to activate the effective ZZ interaction for a specific time of \SI{30}{\nano\second} via the flux controls~\cite{Ren2022Experimental}.
As such, the circuit is recompiled with the native gate set \{$U(\theta, \varphi, \lambda)$, CZ\}, where the arbitrary single-qubit gates are parameterized as
\begin{align}\label{eq:u-gate}
	U(\theta, \varphi, \lambda)=\left( {\begin{matrix} \cos{\frac{\theta}{2}}&-e^{i\lambda}\sin{\frac{\theta}{2}} \\ e^{i\varphi}\sin{\frac{\theta}{2}} & e^{i(\varphi+\lambda)}\cos{\frac{\theta}{2}} \end{matrix}} \right).
\end{align}

\begin{figure}
    \centering
    \includegraphics[width=1\linewidth]{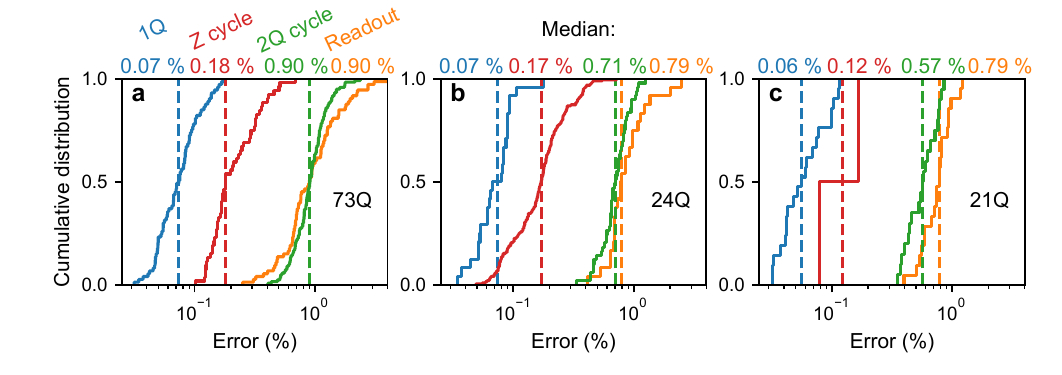}
    \caption{\textbf{Cumulative distributions of errors for simultaneous single-qubit gates (1Q), Z gates (Z cycle, including a Z gate and a single-qubit gate), two-qubit gates (2Q cycle, including a CZ gate and two single-qubit gates) and readout (the average readout error of $\ket{0}$ and $\ket{1}$).} \textbf{a}, $73$ qubits used in the two-dimensional CHSH honeycomb model. \textbf{b}, $24$ qubits used to detect Bell correlation depth. \textbf{c}, $21$ qubits used in the one-dimensional XXZ model.}
    \label{fig:gate_error}
\end{figure}

In our experiment, we achieve high-fidelity parallel quantum gates and projective measurements, which is essential for the demonstration of the efficacy of the VQC-based many-body Bell correlation detector.
For a given quantum circuit, we first reschedule the gate sequences to make them as compact as possible. 
Since the single-qubit gates are realized via microwave controls and the two-qubit gates only involve flux controls, we separate them into different layers to avoid unwanted crosstalk errors. 
Besides, for the two-qubit gate layers, we optimize the work points for every two-qubit gate as well as the idling qubits simultaneously to minimize the errors due to parasitic couplings. 
The performances of the single- and two-qubit gates are assessed with simultaneous cross-entropy benchmark techniques. 
To reduce errors of the readout procedure, we apply a $X_{21}$ pulse, which rotates the transmon in the $\ket{2}\text{--}\ket{1}$ subspace by an angle $\pi$, before applying the readout pulses.
We conduct simultaneous calibration of single- and two-qubit gate error and readout error for each of the three different experiments, the results are shown in Fig.~\ref{fig:gate_error}. 
In the simultaneous two-qubit gate layer, we optimize the energy level of qubits that are not involved in the CZ gates by performing Z gates to avoid interaction with surrounding qubits~\cite{Xu2024Fibonacci}.

\subsection{Many-body quantum correlation measurement}

The detection of many-body quantum correlations is typically challenging in experiments.
In our study, we employ two techniques, namely parity oscillation and multiple quantum coherence, to probe the many-body quantum correlations in our system. 
Below we introduce the two methods and analyze the corresponding advantages and disadvantages in detail.

\subsubsection{Parity oscillation measurement}

The expectation value of the multi-qubit operator $\mathcal{B}_N$~\eqref{eq:bell_operator_N22_Sve+Mer} can be written as
$\langle \mathcal{B}_N \rangle=2^{(N-1)/2}(\langle\mathcal{C}\rangle + \langle\mathcal{C}\rangle^*)=2^{(N+1)/2}\Re\left(\langle\mathcal{C}\rangle\right)$, where the matrix element $\mathcal{C}\equiv |0\rangle\langle 1|^{\otimes N}$, and $\Re\left(x\right)$ denotes the real part of $x$.
Experimentally, $\langle\mathcal{C}\rangle$ can be obtained by parity measurement~\cite{20q_ghz}.
Consider the $N$-qubit parity operator $\mathcal{P}(\gamma)=\otimes_{j=1}^N\left[\sin(\gamma) \sigma_{y, j}+\cos(\gamma) \sigma_{x, j}\right]$, for a given density matrix $\rho$, its expectation value can be expressed as
\begin{align}
    \label{eq:parity}
    \begin{split}
    \langle\mathcal{P}(\gamma)\rangle &=\sum_{\boldsymbol{s}}\rho_{\boldsymbol{s},\boldsymbol{\Bar{s}}} e^{i\left(2\mathbb{Z}_{\boldsymbol{s}}-N\right)\gamma} \\
    &=\sum_{\boldsymbol{s}} \left|\rho_{\boldsymbol{s},\boldsymbol{\Bar{s}}}\right|\cos{\left(\left(2\mathbb{Z}_{\boldsymbol{s}}-N\right)\gamma + \phi_{\boldsymbol{s},\boldsymbol{\Bar{s}}}\right)},
    \end{split}
\end{align}
where $\boldsymbol{s}=s_{1}s_{2}\dots s_{N} (s_{j}\in{\{0,1\}})$ denotes a bitstring and $\boldsymbol{\Bar{s}}$ represents its complement (e.g. $\boldsymbol{s}=00\dots0$ and $\boldsymbol{\Bar{s}}=11\dots1$, hense $\rho_{\boldsymbol{s},\boldsymbol{\Bar{s}}}$ is the anti-diagonal term in the density matrix),
$\phi_{\boldsymbol{s},\boldsymbol{\Bar{s}}}$ is the phase of $\rho_{\boldsymbol{s},\boldsymbol{\Bar{s}}}$,
and $\mathbb{Z}_{\boldsymbol{s}}$ is the number of $0$ in $\boldsymbol{s}$.
By noticing that $\langle\mathcal{C}\rangle=\rho_{00\dots0,11\dots1}$, the correlation can be determined as the Fourier components $\mathcal{P}_f(N), \mathcal{P}_f(-N)$ of $\langle\mathcal{P}(\gamma)\rangle$, where $\mathcal{P}_f(q)=\mathcal{N}_s^{-1}\sum_{\gamma}e^{-iq\gamma}\langle\mathcal{P}(\gamma)\rangle$ is the discrete Fourier transformation and $\mathcal{N}_s$ is the sampling number of $\gamma$.

Experimentally we take a sparse sampling of $\langle\mathcal{P}(\gamma)\rangle$ with $\mathcal{N}_s=N+1$ points, where $\gamma=-\frac{\pi}{2}+\frac{\pi}{N+1}k$ $(k=0,1,\dots,N)$. 
The quantum circuit is illustrated in Extended Data Figure 1(a).
For each $\gamma$ we apply a single-qubit gate $U\left(\pi/2, \gamma-\pi, \pi-\gamma\right)=\frac{\sqrt{2}}{2} \left( {\begin{matrix} 1&e^{-i\gamma} \\ -e^{i\gamma} & 1\end{matrix}} \right)$
to each qubit before simultaneous readout. These rotations bring the axis defined by the operator $\mathcal{P}(\gamma)$ to the $z$ axis. 
Subsequently, we perform simultaneous measurements to all qubits and record the outcomes as binary strings over $M$ shots. From these measurements, we compute the parity expectation $\left\langle \mathcal{P}(\gamma) \right\rangle=\frac{M_{+} - M_{-}}{M}$, where $M_{+}$ ($M_{-}$) represents the count of occurrences corresponding to eigenvalue $1$ ($-1$). 
For each $\left\langle \mathcal{P}(\gamma)\right\rangle$, we take $900$, $1500$, $2400$, $3600$, $5000$, $7200$, $9600$, $12000$, $15000$, $20000$, $25000$, $30000$ measurement shots for qubit number $N=2,4,6,\dots,24$.
To correct the readout error, we use the tensor product of single-qubit correction matrices.

\begin{figure}
    \centering
    \includegraphics[width=1\linewidth]{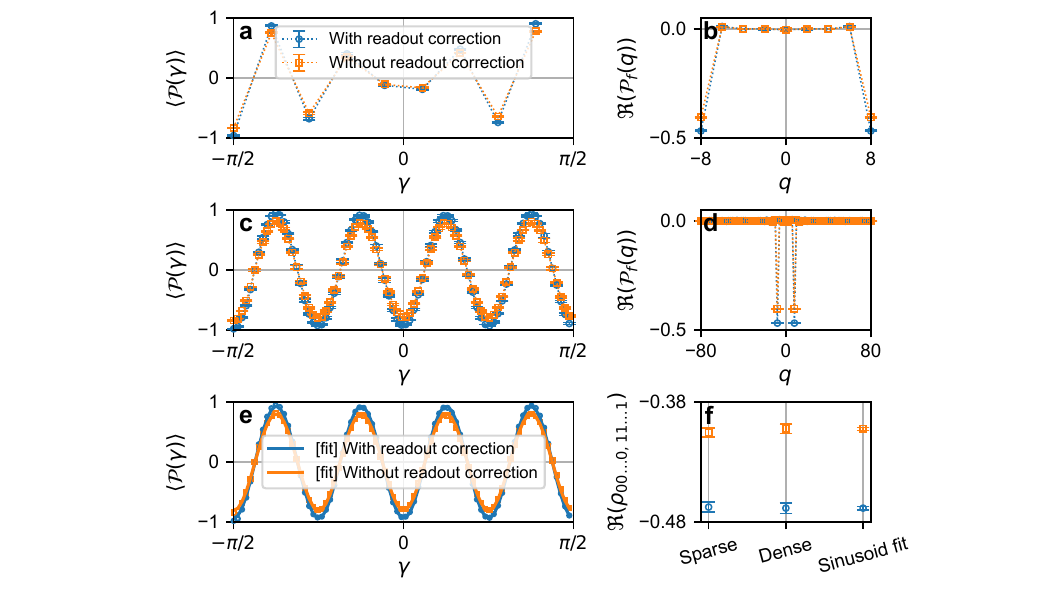}
    \caption{\textbf{Contrast three different methods for analyzing $8$-qubit parity oscillation measurement results: sparse sampling, dense sampling, and sinusoid fitting.} \textbf{a} and \textbf{b}, Sparse sampling with sampling size $\mathcal{N}_s=9$ and the real part of its Fourier spectra. The Fourier component $\Re\left(\mathcal{P}_f(8)\right)$ equals to $\Re\left(\rho_{00\dots0,11\dots1}\right)$. \textbf{c} and \textbf{d}, Dense sampling with sampling size $\mathcal{N}_s=81$ and the real part of its fourier spectra. \textbf{e}, Sinusoid fit applied to the dense sampling data, with the frequency set to $8$. \textbf{f}, Comparison of the outcomes obtained from the three different methods.}
    \label{fig:compare_dense_sparse}
\end{figure}

In order to test the accuracy and reliability of this sparse sampling method, we compare the outcomes of sparse sampling ($\mathcal{N}_s=N+1$) with dense sampling ($\mathcal{N}_s=10N+1$) case for an experimentally trained 8-qubit GHZ state. In the dense sampling case, we apply both the Fourier transformation and sinusoid fitting to the experimental data, with the results shown in Fig.~\ref{fig:compare_dense_sparse}. The extracted correlations are similar across all three cases.

\subsubsection{Multiple quantum coherence}

The parity oscillation (PO) measurement offers an efficient way to detect the many-body correlation in this work.
Now we introduce another technique, i.e., the Multiple Quantum Coherence (MQC)~\cite{MQC2020}, which can also measure the many-body correlation efficiently.
In addition, the MQC method is less susceptible to readout errors, as we will elaborate in the next section. 

The quantum circuit for MQC is depicted in Extended Data Figure 1(b). 
Following the variationally obtained $N$-qubit state $\rho=U_{\text{exp}}\ket{00\dots0}\bra{00\dots0}U_{\text{exp}}^\dagger$, we introduce an $X$ gate $U_X$ and a phase gate $U_\phi=e^{-i\frac{\phi}{2}\sigma_z}=\left( {\begin{matrix} e^{-i\phi/2} & 0 \\ 0 & e^{i\phi/2}\end{matrix}} \right)$ to each qubit, succeeded by a reversal circuit $U^{\dagger}_{\text{exp}}$.
After all these operations, the probability of detecting the ground state becomes

\begin{equation}
\begin{aligned}
\mathcal{K}(\phi)
&= \left|\bra{00 \dots0} U_{\text {exp }}^{\dagger} U_{\phi}^{\otimes N} U_{X}^{\otimes N} U_{\text {exp }} \ket{00 \dots0}\right|^2 \\ 
& =\sum_{\boldsymbol{s},\boldsymbol{s'}}\rho_{\boldsymbol{s},\boldsymbol{s'}} \rho_{\boldsymbol{\Bar{s'}}, \boldsymbol{\Bar{s}}} e^{i\left(\mathbb{Z}_{\boldsymbol{s}} - \mathbb{Z}_{\boldsymbol{s'}}\right)\phi} 
\end{aligned}
\end{equation}
where $\boldsymbol{s}$($\boldsymbol{s'}$) denotes a bitstring and $\boldsymbol{\Bar{s}}$($\boldsymbol{\Bar{s'}}$) represents its complement, $\mathbb{Z}_{\boldsymbol{s}}$($\mathbb{Z}_{\boldsymbol{s'}}$) is the number of 0 in $\boldsymbol{s}$($\boldsymbol{s'}$).

Noting that the oscillation terms with the maximum frequency $N$ ($-N$) in $\mathcal{K}(\phi)$ only occur for $\boldsymbol{s}=00\dots0$ ($\boldsymbol{s}=11\dots1$) and $\boldsymbol{s'}=11\dots1$ ($\boldsymbol{s'}=00\dots0$), and that $\rho_{00\dots0,11\dots1}=\rho^*_{11\dots1, 00\dots0}$, we can obtain $|\rho_{00\dots0,11\dots1}|$ by Fourier transforming $\mathcal{K}(\phi)$ at the frequency $N$ as
\begin{align}
    \label{eq:MQC}
    \begin{split}
|\rho_{00\dots0,11\dots1}|
= \frac{1}{\sqrt{\mathcal{N}_s}}\sqrt{\left|\sum_\phi e^{-iN\phi} \mathcal{K}(\phi)\right|},
    \end{split}
\end{align}
where the normalization constant $\mathcal{N}_s$ represents the sampling number of $\phi$.
The phase of $\rho_{00\dots0,11\dots1}$ cannot be directly obtained due to potential imperfections in the reversal circuit.
Therefore, we rely on the phase extracted from the parity oscillation measurement to calculate the real part of $\rho_{00\dots0,11\dots1}$.

Experimentally we only employ MQC as a means to detect genuine multipartite Bell correlations on the variationally trained state. To accurately measure $\mathcal{K}_f(N)$ without compromising efficiency, we adopt sparse sampling on $\phi$, where $\phi=\frac{\pi j}{N+1}$ for $j=0,1,2,\dots,2N+1$ (Note that the number of sampling points is approximately double that of the parity oscillation measurement.). For each $\mathcal{K}(\phi)$, we employ an equal number of measurement shots as utilized in the parity measurement. Moreover, we implement the same readout correction procedure as employed in the parity measurement.

\subsubsection{Effect of readout error}

As mentioned earlier, the parity oscillation measurement is more susceptible to readout errors. This is primarily due to two reasons. First, the parity is calculated from the entire set of $2^N$ probabilities, whereas MQC only requires measuring the probability of the ground state $|00...0\rangle$. In addition, the MQC circuit involves a reversal circuit, which mitigates the impact of readout errors by taking the square root of the Fourier outcomes.

The projective quantum state readout procedure in superconducting circuits is error-prone, meaning that we may assign a wrong state $|1\rangle$ ($|0\rangle$) to the qubit while it is actually in $|0\rangle$ ($|1\rangle$) with the probability $e_0$ ($e_1$).
To understand the effect of readout error, let's first assume that $e_0=e_1=e$.
As a reference, the experimentally calibrated median values of $e_0$ and $e_1$ are $0.8\%$ and $0.9\%$, respectively.
For the MQC method, if the variational circuit $U_{\text{exp}}(\boldsymbol{\theta})$ perfectly generates a GHZ state, then the final state should be $\frac{1}{2} \left[\left(1+e^{-iN\phi}\right)\ket{0}+\left(1-e^{-iN\phi}\right)\ket{1}\right]\otimes\ket{00\dots0}$, where only two bitstrings, i.e., $00\dots0$ and $10\dots0$, appear.
In the presence of readout errors, the measured ground state probability is given by:
\begin{equation}
\label{eq:read_error_MQC}
\begin{split}
\mathcal{K}(\phi)_{\text{meas}}&=\left(1-e\right)^{N}P_{00\dots0}+e\left(1-e\right)^{N-1}P_{10\dots0} \\ &\approx\left(1-e\right)^{N}\mathcal{K}(\phi),
\end{split}
\end{equation}
where $P_{\boldsymbol{s}}$ denotes the ideal probability of measuring the bitstring state $|\boldsymbol{s}\rangle$. 
The approximation arises because $e\left(1-e\right)^{N-1}P_{10\dots0}$ is much smaller than $1$ due to $e\ll1$, and therefore it can be ignored.
In the experiment, there may be other measured bitstrings besides $00\dots0$ and $10\dots0$ due to imperfections of the quantum gate or finite coherence time. 
Nonetheless, their occurrences are significantly less frequent. Combined with Eq.~\ref{eq:MQC}, we see that due to readout error, the experimentally obtained correlation $|\rho_{00\dots0,11\dots1}|$ is reduced by a factor of $\left(1-e\right)^{N/2}$.

On the other hand, the parity expectations are calculated from the entire set of $2^N$ probabilities, which can be separated by their parity eigenvalue $1$ ($-1$) into two groups.
Let's assume we take $M$ measurement shots, and the count of occurrences corresponding to eigenvalue $1$ ($-1$) is $M_{+}$ ($M_{-}$). 
In the presence of readout errors, the experimentally measured parity expectation can be calculated as:
\begin{align}
\begin{split}
\label{eq:read_error_parity}
\left\langle \mathcal{P}(\gamma)\right\rangle_{\text{meas}}&=\frac{\left[\left(1-e\right)^{N}M_+ + C^1_N e\left(1-e\right)^{N-1}M_- + C^2_N e^2\left(1-e\right)^{N-2}M_+ + \dots\right]}{M} \\ &\ \ \ \ \ \ -\frac{\left[\left(1-e\right)^{N}M_- + C^1_N e\left(1-e\right)^{N-1}M_+ + C^2_N e^2\left(1-e\right)^{N-2}M_- + \dots\right]}{M} \\ &=\Sigma_{i=0}^{N} C_N^i \left(-e\right)^{i} \left(1-e\right)^{N-i} \frac{M_+ - M_-}{M}\\ &=\left(1-2e\right)^{N}\left\langle \mathcal{P}(\gamma)\right\rangle
\end{split}
\end{align}
The first equality is based on the observation that the sign of eigenvalue will change if an odd number of qubits experience a bit-flip readout error.

Finally, the ratio of outcomes between MQC and parity measurement is approximately $\left(1-e\right)^{N/2}/\left(1-2e\right)^{N}$. We compare this estimation with the experimental results for different qubit numbers, as shown in Extended Data Fig.~3.

\subsubsection{Statistical analysis}

For the measured many-body quantum correlation to carry statistical meaning, it needs to be supplemented with an estimate of its standard error. 
A smaller standard error indicates higher confidence in the measurement results and reduces the number of shots required to assert genuine multipartite Bell correlations.
Experimentally, we observe a ratio of approximately $1$:$4$ between the standard errors from MQC and parity measurements of the measured many-body quantum correlation. 
This observation can be attributed to three main factors:

First, the parity expectation is calculated from $\left\langle \mathcal{P}(\gamma)\right\rangle=\frac{M_+ - M_-}{M}=\frac{2 M_+}{M} - 1$ where $M_+$ follows a binomial distribution, whereas in the MQC measurement the outcomes are obtained directly from the ground state probability, which also follows a binomial distribution.
The standard deviation of the parity expectation is given by $\sigma\left(\left\langle \mathcal{P}(\gamma)\right\rangle\right)=2\sqrt{\frac{p\left(1-p\right)}{M}}$ where $p$ is the probability of getting eigenvalue $1$ in a single shot. 
If we assume in the MQC measurement the probability of observing the ground state is also $p$ and take the same number of measurement shots, the standard deviation of the MQC outcome is only $\sqrt{\frac{p\left(1-p\right)}{M}}$.

Second, the number of sampling points in the discrete Fourier transformation of MQC is approximately double that of the parity oscillation measurement, which reduces the standard deviation by a ratio of $\sqrt{2}$. 
However, it's important to note that this reduction in standard deviation is actually due to the total increase in the number of measurement shots taken by the MQC measurement, rather than being an inherent advantage of the MQC technique itself.

Third, observation from Eq.~\eqref{eq:read_error_MQC} shows that the readout error changes the range of ground state probability from symmetry around $0.5$ to below it (the standard deviation of a binomial distribution is maximal when $p=0.5$). 
In contrast, in the parity oscillation measurement, the readout error causes the parity expectation to remain symmetric around $0$ (the standard deviation is maximal when the parity expectation is $0$).
Therefore, the MQC benefits from this asymmetry induced by the readout error, leading to a further reduction in the standard deviation.

\bibliography{QMLBib}